\documentclass{aa}

\usepackage{graphicx}
\usepackage{txfonts}

\usepackage{amstext}
\usepackage{ulem}

\usepackage[colorlinks=true,citecolor=blue]{hyperref}
\usepackage{xspace}
\usepackage{mathtools}

\newcommand{\LEt}[1]{}
\renewcommand{\bf}[1]{#1}
\renewcommand{\sout}[1]{}

\titlerunning{Simulated LSB clusters}
\authorrunning{Ragagnin et al.}

\begin{document}

   \title{%Gas-poor galaxy clusters in  cosmological simulations 
   %\sa{
   %Why some galaxy clusters are gas-poor? A simulation view.
   %}
   Simulation view of galaxy clusters with low X-ray surface brightness
   }

   \author{A. Ragagnin\inst{1,2,3},
           S. Andreon\inst{4},
           E. Puddu\inst{5}
          }

   \institute{Dipartimento di Fisica e Astronomia "Augusto Righi", Alma Mater Studiorum Università di Bologna, via Gobetti 93/2, I-40129 Bologna, Italy\\
            \email{antonio.ragagnin@unibo.it}
            \and
            INAF - Osservatorio Astronomico di Trieste, via G.B. Tiepolo 11, 34143 Trieste, Italy
            \and
            IFPU - Institute for Fundamental Physics of the Universe, Via Beirut 2, 34014 Trieste, Italy
            \and
            INAF - Osservatorio Astronomico di Brera, Via Brera 28, 20121 Milano, Italy
             \and
             INAF - Osservatorio di Capodimonte, Salita Moiariello 13, 80131 Napoli, Italy
             }

   \date{accepted}

% \abstract{}{}{}{}{} 
% 5 {} token are mandatory
 
  \abstract
  % context heading (optional)
{
X-ray\LEt{please choose a running title that consists of actual words} selected samples are known to miss galaxy clusters that are gas poor and have a low surface brightness. This is different for the optically selected samples  such as the X-ray Unbiased Selected Sample (XUCS).
}   
{We characterise  the origin of galaxy clusters that are gas poor and have a low surface-brightness\LEt{you just introduced an abbreviation for this. Please decide whether you wish to use this or not and then change accordingly in the Abstract. The Abstract is considered a stand-alone item, therefore all abbreviations that you wish to use need to be introduced again and are then to be used throughout in the main text. Please check this and change as required, I'll not highlight this again to avoid cluttering the ms} by studying covariances between various
cluster properties at fixed mass using hydrodynamic cosmological simulations.}
{We extracted $\approx 1800$ galaxy clusters from a high-resolution Magneticum hydrodynamic cosmological simulation
and computed covariances at fixed mass of the following properties: core-excised X-ray luminosity, gas fraction,  hot gas temperature, formation redshift, \LEt{pleas spell this out, you only use it this once in the Abstract}{\bf matter density profile} concentration, galaxy richness, fossilness parameter, and stellar mass of the bright central galaxy. We also compared the correlation between concentration 
and gas fractions in non-radiative simulations, and we followed the trajectories of particles inside galaxy clusters to assess the role of AGN depletion on the gas fraction.}
{In simulations and in observational data, differences in surface brightness are related to differences in
gas fraction. Simulations show that the gas fraction strongly correlates with assembly time, in the sense that older clusters are gas poor.
Clusters that formed earlier have lower gas fractions
because the feedback  of the active galactic nucleus
ejected a significant amount of gas from the halo.
When the X-ray luminosity is corrected for the 
gas fraction, 
  it shows little or no covariance with other quantities.
}
   {Older galaxy clusters tend to be gas poor and possess a low  X-ray  surface brightness because the feedback mechanism  removes a significant fraction of gas from these objects. Moreover, 
    we found that most of the $L_X$ covariance with the other quantities is explained by differences in the gas fraction.}


   \maketitle

%-------------------------------------------------------------------

\section{Introduction}

Galaxy clusters \citep[see][for a review]{2012ARA&A..50..353Kravtsov}
are the most massive gravitationally bounded structures of our Universe.
Their masses cannot be observed directly, unless through weak-lensing observations, which require a number of assumptions and high-quality data, however. This means that precise estimations are rare~\citep{2010ApJ...721..875Okabe,2012MNRAS.427.1298Hoekstra,2015MNRAS.449.2219Melchior,2019MNRAS.485...69Stern}. To estimate galaxy cluster masses,  it is important to calibrate mass-observable relations ~\citep{2013SSRv..177..247Giodini,2011ARA&A..49..409Allen,2021MNRAS.505.3923Schrabback} and their scatter values~\citep{2005PhRvD..72d3006Lima}.
For this reason, it is important to estimate how the scaling relations are affected by processes such as the environment and accretion history~\citep{2018ARA&A..56..435Wechsler}, cosmological parameters~\citep{2021ApJ...911...82Corasaniti}, and the theoretical modelling of baryon physics~\citep{2005ApJ...634..964Ostriker,2006MNRAS.365.1021Ettori,2007MNRAS.377...41Crain,2009ApJ...700..989Bode,2015MNRAS.450..896Dvorkin,2020NatAs...4...10Gaspari,2021MNRAS.500.2316Castro,2021ApJ...921..112BeltzMohrmann,2021JCAP...03..023Khoraminezhad}.

This estimation is complicated by the fact that X-ray luminosity ($L_X$) observations may be  biased towards measurements of higher $L_X$ and centrally peaked gas distributions~\citep{2007MNRAS.382.1289Pacaud,2010A&A...513A..37Hudson,2011A&A...536A..37Andreon,2016A&A...593A...2Andreon,2018A&A...619A.162Xu},
which may imply that these surveys are biased towards galaxy clusters that are more gas rich and X-ray bright.
For this reason, it is useful to characterise the population of galaxy clusters with low X-ray surface brightness (LSB) that are gas poor.
The X-ray Unbiased Selected Sample \citep[XUCS;][]{2016A&A...593A...2Andreon} contains
galaxy clusters that at fixed mass are 
gas poor, have an LSB \citep{2017A&A...606A..24Andreon},
posses lower richness  \citep{2022MNRAS.511.2968Puddu}
at fixed mass, and do not show an indication of an anti-correlation between cold baryons  (e.g. richness), and hot baryons (such as gas mass), as has been suggested in ~\cite{2019NatCo..10.2504Farahi}.

Recent numerical studies hint a positive correlation between cold and hot baryons as richness at fixed mass anti-correlates with concentration~\citep{2019MNRAS.490.5693Bose}, which itself is an index of the dynamical state~\citep{2012MNRAS.427.1322Ludlow}, and un-relaxed systems tend to be gas rich ~\citep{2020MNRAS.491.4462Davies}.
However, these studies were carried out over different simulation suites, while the more comprehensive numerical work of  \cite{2010ApJ...715.1508Stanek} shows a positive correlation between gas fraction and concentration.
One possible explanation  of this mismatch might be their different baryon physics implementation.

For this reason, we conducted a comprehensive study of LSB and gas-poor clusters in modern cosmological simulations including all main baryon physics processes, such  as  stellar evolution and  active galactic nucleus (AGN) feedback (hereafter full-physics), which reproduce reasonable scaling relations~\citep{2018MNRAS.474.4089Truong} and optical  properties~\citep{2020MNRAS.495..686Anbajagane}.  
Given these premises and because observational data have no access to the accretion history of  galaxy clusters (if not its imprints, such as the concentration), we used full-physics cosmological simulations  to characterise  LSB and X-ray bright galaxy clusters by studying their  gas fraction  and concentration.
We assessed the role of baryon physics by inter-connecting these properties by comparing full-physics and non-radiative simulations (thus without star formation).
In particular,  we employed Magneticum\footnote{\url{http://www.magneticum.org}} hydrodynamic simulations \citep{2013MNRAS.428.1395Biffi,2014MNRAS.440.2610Saro,2015MNRAS.448.1504Steinborn,2016MNRAS.463.1797Dolag,2015MNRAS.451.4277Dolag, 2015ApJ...812...29Teklu,2016MNRAS.458.1013Steinborn,2016MNRAS.456.2361Bocquet,2019MNRAS.486.4001Ragagnin}  that proved to produce realistic galaxies and clusters~\citep[see e.g.][]{2015ApJ...812...29Teklu,2019MNRAS.484..869VanDeSande,2017MNRAS.464.3742Remus}.

When we compared simulations and observations, we used similar cosmologies. We used the  WMAP7 cosmology~\citep{2011ApJS..192...18Komatsu} for simulations, but observations assume  $\Omega_m=0.3$ and  $h_0=0.7.$

In Section \ref{sec:data} we present the simulation and observational data.    In Section \ref{sec:lxm} we show the effect of gas-mass correction in the $L_X-M$ relation. We discuss our results in Section \ref{sec:discu} and draw our conclusions in Sec. \ref{sec:conclu}.

\section{Data}
\label{sec:data}

\begin{figure*}
\centerline{
\includegraphics[width=9truecm]{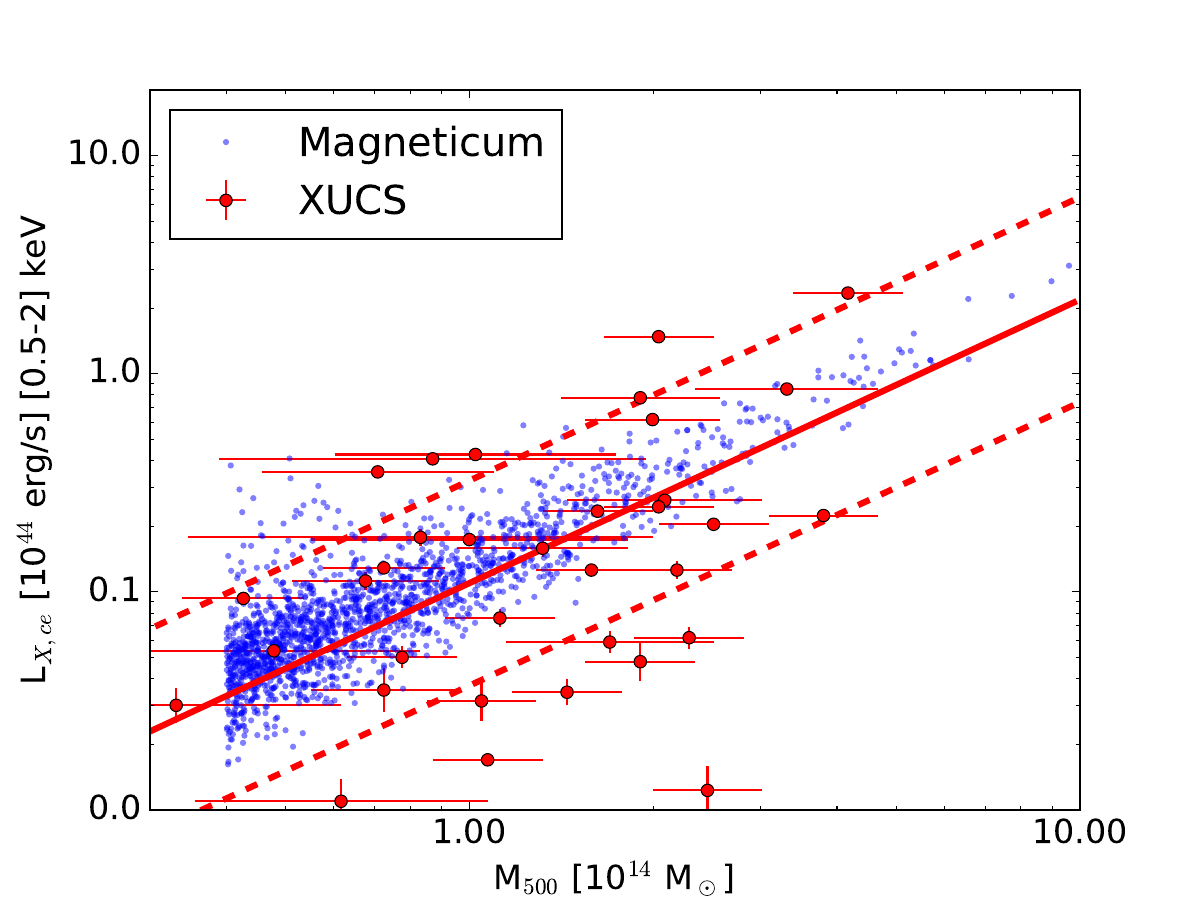}%
\includegraphics[width=9truecm]{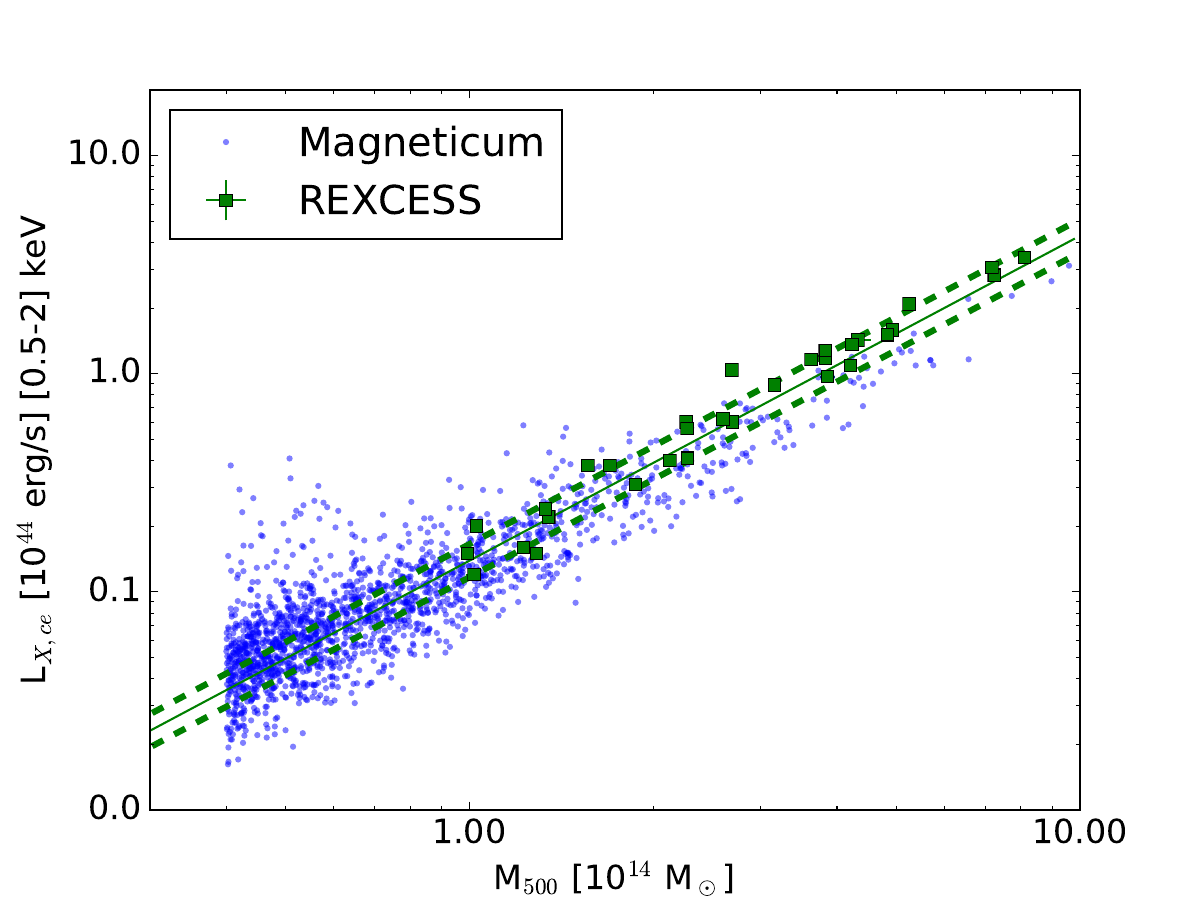}%
}
\caption[h]{X-ray luminosity in an X-ray unbiased sample (red points, left panel) in the X-ray selected REXCESS sample (green points, right panel) and in the Magneticum simulation (blue points, both panels). Best fit and  $\pm 1\sigma$ intrinsic scatter corridors are also shown for the observed data. Photometric errors boost the scatter in the observational  data.
}
\label{fig:LxM}
\end{figure*}

\subsection{Numerical setup}

The Magneticum simulations are based on the  N-body code \texttt{Gadget3}, which is an improved version of \texttt{Gadget2} \citep{2005Natur.435..629Springel,2005MNRAS.364.1105Springel,2009MNRAS.398.1150Boylan}, with a space-filling curve-aware neighbour search \citep{2016pcre.conf..411Ragagnin} and the improved smoothed particle hydrodynamics (SPH)  solver from \cite{2016MNRAS.455.2110Beck}.
These simulations include a treatment of radiative cooling, heating, ultraviolet (UV) background,  star formation, and stellar feedback processes as in \cite{2005MNRAS.361..776Springel}, connected to a detailed chemical evolution and enrichment model as in \cite{2007MNRAS.382.1050Tornatore}, which follows  11 chemical  elements (H, He,
C, N, O, Ne, Mg, Si, S, Ca, and Fe) with the aid of CLOUDY photo-ionisation
code \citep{1998PASP..110..761Ferland}.  \cite{2010MNRAS.401.1670Fabjan,2014MNRAS.442.2304Hirschmann} described prescriptions for black hole growth and feedback from AGNs.
Haloes and their member galaxies are identified using the \LEt{please introduce} friend-of-friend halo finder~\citep{1985ApJ...292..371Davis} and  an improved version of the subhalo finder  SUBFIND \citep{2001MNRAS.328..726Springel},   which takes the presence of baryons into account \citep{2009MNRAS.399..497Dolag}.
We mainly focus on   Magneticum Box2/hr\footnote{Box2/hr halo data are available in the web portal presented in \cite{2017A&C....20...52Ragagnin}}~\citep{2014MNRAS.442.2304Hirschmann}, covering a length of $\approx 900$ comoving Mpc,  with dark matter particle masses $m_\mathrm{DM } = 9.8\cdot10^8M_\odot$ and gas initial particle masses of $m_\mathrm{gas} = 2\cdot10^8M_\odot$ and a gravitational softening of both gas and dark matter of  $\epsilon \approx 5$ comoving kpc\LEt{a single sentence does not constitute a paragraph. Please either add to this or merge}.

Throughout this paper, we use $R_{\rm 500c}$  and $R_{\rm 200c}$ as halo radii. They are defined to be the radius at which the average density of a galaxy cluster
reaches $500$ or $200$ times the critical density of the Universe; see \cite{2015MNRAS.447.1873Naderi} for a review of computing masses and radii within a given overdensity.

We extracted all Box2/hr galaxy clusters with $M_{\rm 500c}$  (i.e. the total mass within $R_{\rm 500c}$) greater than $4\times10^{13}{\rm M}_\odot$ at a redshift $z=0$ for a total of $\approx 1800$ galaxy clusters. For each halo, we computed the following quantities\LEt{please reformat this so that it becomes a proper paragraph. There is no need for the bullet points}:
the gas fraction $f_g,$ from which we removed   star-forming particles by considering only particles with $T>3\times10^{4}K$ and with a cold-gas fraction greater than $0.1;$  
the formation redshift $z_f,$   at which the halo accreted $50\%$ of its final virial mass; The halo concentration $c$,  computed by fitting an NFW~\citep{1997ApJ...490..493Navarro}  profile of the average total (i.e. including dark matter and baryons) matter profile over $40$ logarithmically spaced radial bins in the $[0.01-1]R_{\rm {500c}}$ range; 
    the hot-gas temperature $T,$ computed by mass-weighting SPH particle temperatures with the same filter as the gas fraction;
    the fossilness  within $R_{500c},$ which is the stellar mass ratio of the BCG and the most massive satellite~\citep{2019MNRAS.486.4001Ragagnin}; 
   the richness $n$ of all subhaloes with stellar mass $M_\star>3\times10^{10}M_\odot;$
   the BCG stellar mass $M_{\star, {\rm BCG}},$ as provided by SUBFIND;
     The core-excised X-ray luminosities $L_{X, {\rm ce}}$ were computed using the APEC model \citep{2001ApJ...556L..91Smith}, which considers the emission of a collisionally ionised, chemically enriched plasma implemented within the external, publicly
available package XSPEC\footnote{\url{https://heasarc.gsfc.nasa.gov/xanadu/xspec/}}~\citep{1996ASPC..101...17Arnaud} in the [0.15-1] $R_{\rm 500c}$ range,
 in the $\left[0.5,2\right]$ keV band. We opted for an $L_x$ that was core-excised in order to minimise the impact of details of the sub-grid AGN feedback and of a cool core.

Although fossilness, richness, and BCG mass should correlate with concentration, we decided to include all these indexes in our study because there is no general agreement about the cold-hot baryon correlation in the literature. In this work, all masses and radii are expressed in physical units. They are therefore not implicitly divided by $(1+z)$ or $h_0$ as in other works on simulations.

\begin{figure*}
\centerline{
\includegraphics[width=9truecm]{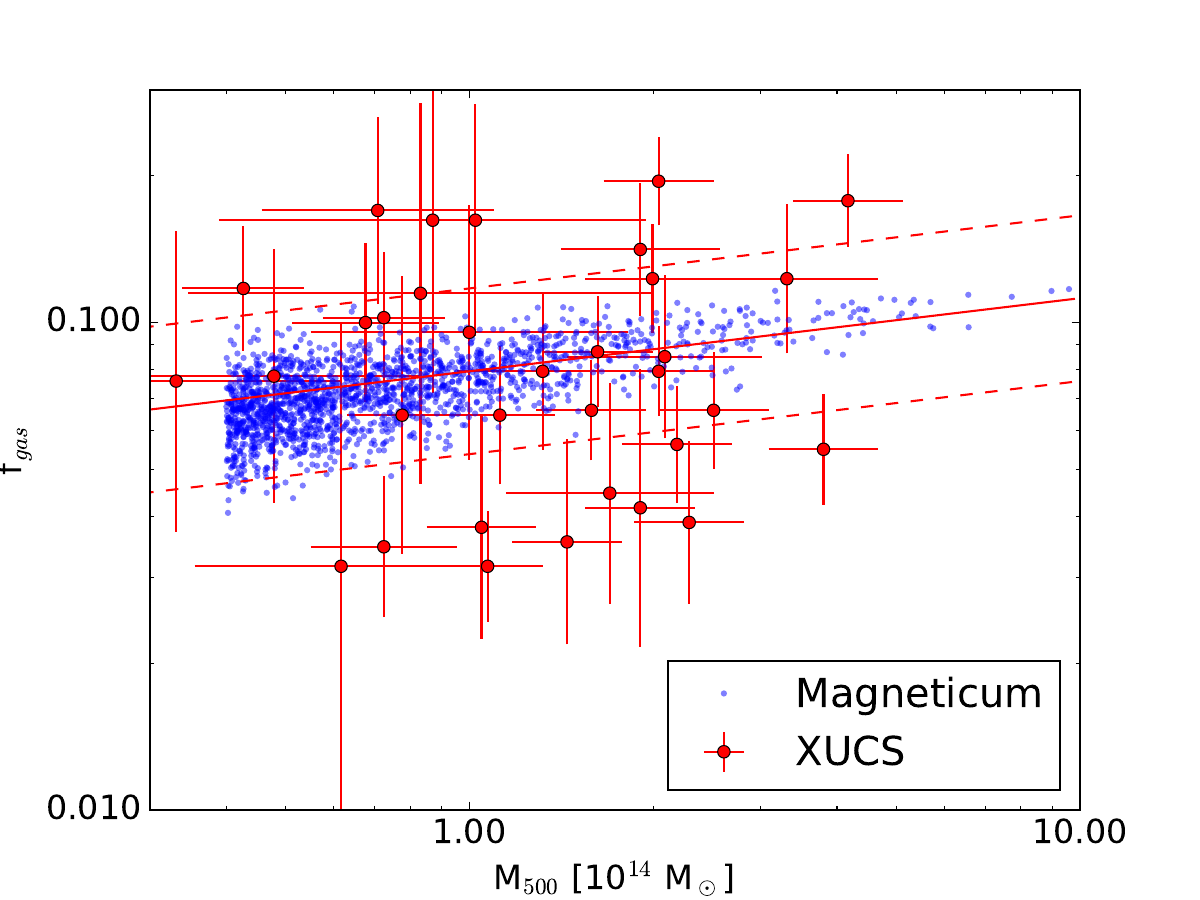}%
\includegraphics[width=9truecm]{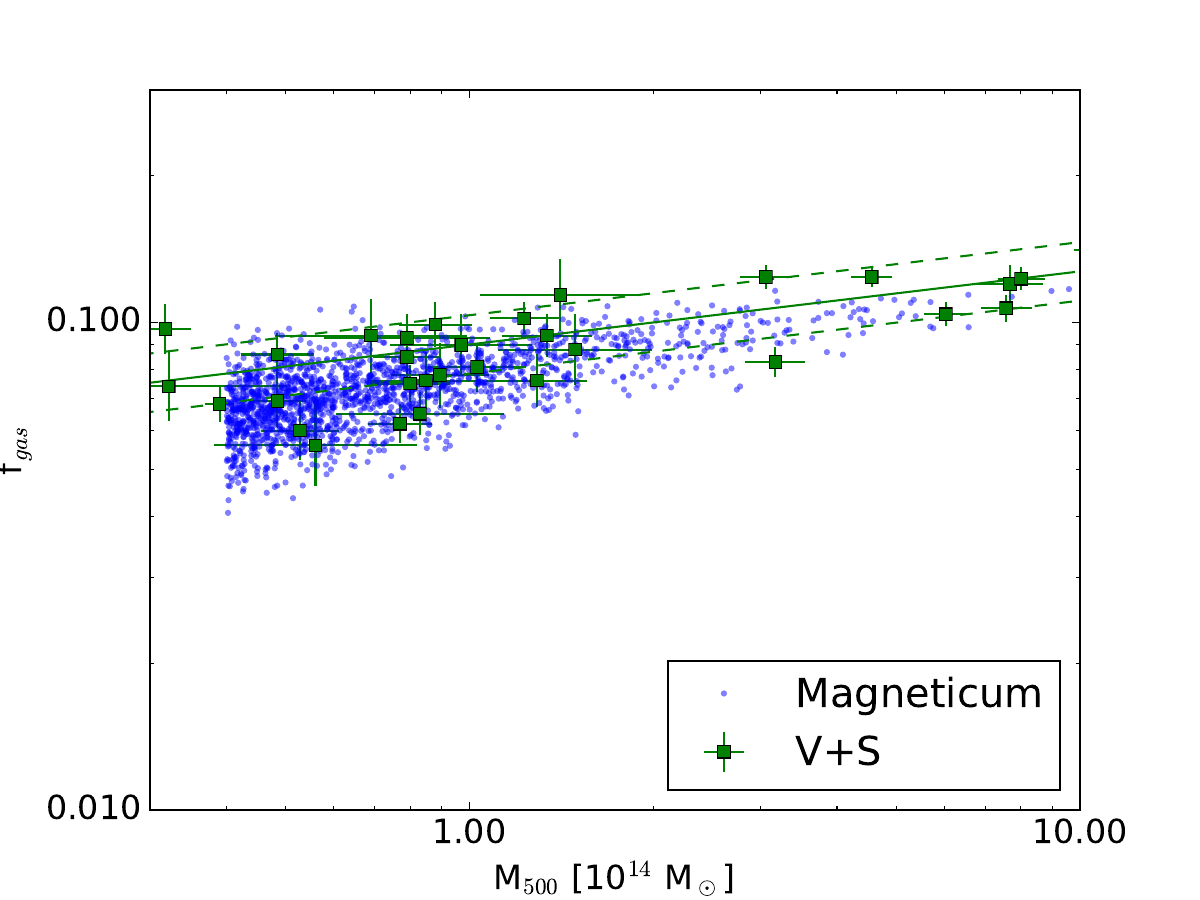}%
}
\caption[h]{Gas fraction in an X-ray unbiased sample (red points, left panel), in the X-ray selected sample (green points, right panel), and in the Magneticum simulation (blue points, both panels). Best fit and $\pm 1\sigma$ intrinsic scatter corridors are also shown. Photometric errors of the observational samples boost the data scatter.
}
\label{fig:fgasM}
\end{figure*}

\subsection{Observational data}
\label{sec:od}
We compared simulated data against the XUCS and to two X-ray selected samples,
REXCESS and an archive sample.
XUCS (Paper I) consists of 34 clusters in the very
nearby universe ($0.050 < z < 0.135$) selected from
the SDSS spectroscopic survey using more than 50
concordant redshifts whithin 1 Mpc and a velocity dispersion of
members $\sigma_v>500$ km/s \citep[see Paper I and][hereafter Paper II]{2017A&A...606A..24Andreon}. They are by design in regions 
of low galactic absorption. 
There is no X-ray selection in the sample in the sense that the
probability of including a cluster is independent of
its X-ray luminosity, or any X-ray property, and no cluster is added or removed on the basis of its
X-ray luminosity. All clusters were followed up in X-ray with Swift,
except for a few with adequate XMM-Newton or Chandra data in the archives. From these
data, we derived the X-ray luminosity with a mean error of 0.04 dex \citep{2016A&A...593A...2Andreon} and the gas fraction
with a mean error of 0.10 dex \citep{2017A&A...606A..24Andreon}. The masses were obtained with the 
caustic technique \citep[and later works]{1997ApJ...481..633Diaferio}, which has the advantage of
not relying on the dynamical equilibrium hypothesis. The masses have
an average mass error of 0.14 dex. In XUCS, caustic masses are consistent with dynamical and weak-lensing
masses (\citealt{2016A&A...593A...2Andreon} and \citealt{2017A&A...606A..25Andreon}),
and for a single cluster with a deep X-ray follow-up, also with
the hydrostatic mass ~\citep{2019A&A...630A..78Andreon}. In other samples, caustic masses are in general found to
agree with weak-lensing masses and hydrostatic masses \citep{2013ApJ...764...58Geller,2015MNRAS.449..685Hoekstra,2016MNRAS.461.4182Maughan}.

REXCESS \citep{2007A&A...469..363Boehringer} consists of an X-ray selected sample of 33 clusters at low redshift ($z<0.2$)
selected from the ROSAT all-sky survey \citep{1993Sci...260.1769Truemper} that was followed up with XMM-Newton. Cluster masses were derived from the Compton $Y_X$ parameter
\citep{2010A&A...517A..92Arnaud}, and X-ray luminosities were derived in \cite{2009A&A...498..361Pratt}. This sample lacks gas-fraction
determinations, and therefore we supplemented it with measurements from
\cite{2006ApJ...640..691Vikhlinin} and  \cite{2009ApJ...693.1142Sun} (V+S, hereafter). The latter sample was observed with Chandra, has $z<0.25, $ and is formed by 30 clusters 
and groups, 3 of which are present in both papers, with measurements at
$R_{\rm 500c}$. This sample lacks a selection function and uses hydrostatic masses.

\begin{figure}
\centerline{\includegraphics[width=.9\linewidth]{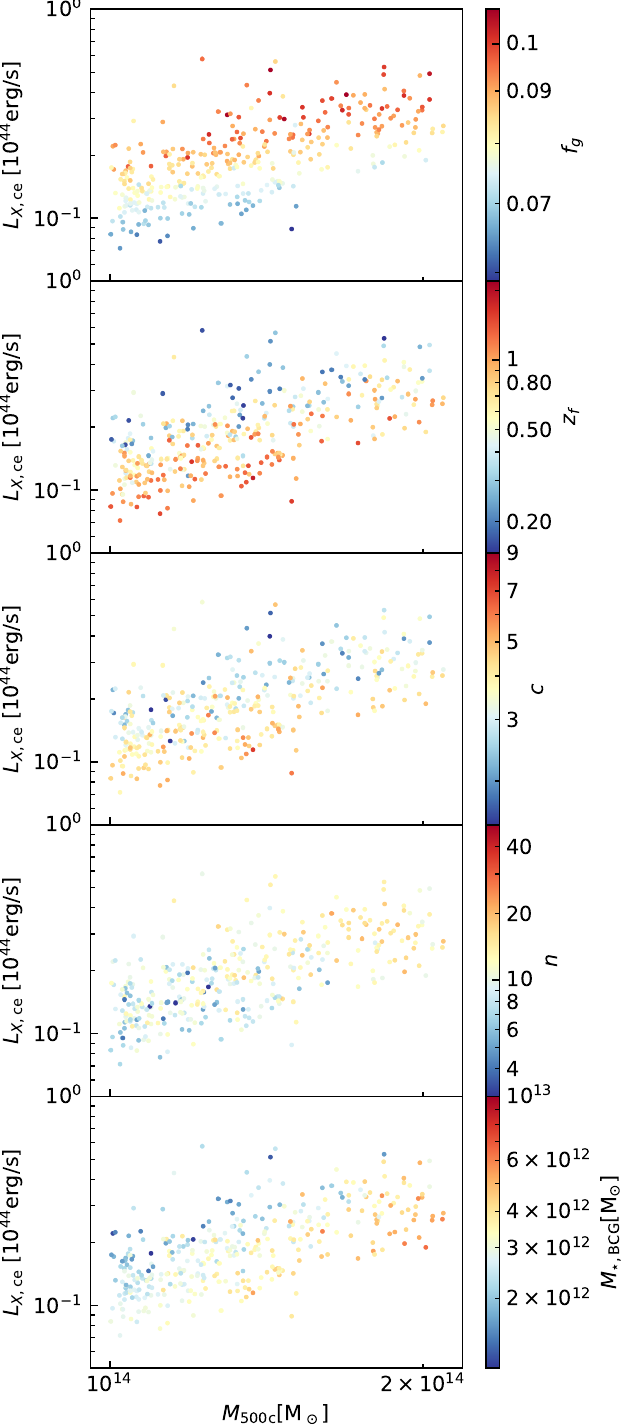}}
\caption[h]{Core-excised X-ray luminosity $L_{X, \rm ce}$ $vs.$ mass $M$ in
the Magneticum simulation colour-coded by gas-fraction (top panel), formation epoch $z_f$ (second panel from the top),  concentration $c$ (central panel), richness $n$ (second panel from the bottom), and BCG stellar mass (bottom panel).
To illustrate the dependence at fixed mass,
we only show data in a small halo mass range ($1-2\times10^{14}{\rm M}_\odot$).
\LEt{a figure caption should only contain the information required to understand the figure, no interpretation or conclusions. To my understanding, the next sentence should be removed}
}
\label{fig:LxMcolor}
\end{figure}

\section{Luminosity-mass relation}
\label{sec:lxm}

In this  section we compare the $L_X-M$ relation of simulations against the biased and unbiased X-ray samples.
Figure \ref{fig:LxM} shows the relation of $L_{X,ce}$ versus $M_{500c}$  for simulated haloes overplotted on the observed X-ray selected (right panel) and X-ray unbiased (left panel) samples. 
Simulations and real data share similar slopes (Magneticum: 1.23, XUCS:1.30, and REXCESS: 1.49). The scatter at fixed mass of simulated clusters
is between 0.24 dex (at the high-mass end) to 0.30 dex (in the low-mass range), which is smaller than XUCS (0.47 dex), but much larger than REXCESS (0.08 dex). 

The Magneticum feedback parameters were calibrated to  match observed quantities such as cluster baryon-fractions ~\citep{2016MNRAS.456.2361Bocquet}  within the uncertainty of the observational data. We therefore do not focus on the exact value of scaling-relation normalisation values, but rather on the scatter at fixed mass.

The Magneticum mean relation has a lower intercept than that of REXCESS because the latter misses clusters with low gas-fractions. The relation
is higher than XUCS, however, indicating that gas-poor clusters are more abundant in the Universe than in Magneticum.
The larger intercept and lower scatter of X-ray selected samples are known features of samples selected in this way~\citep{2016A&A...593A...2Andreon}.
While simulations do not capture the full variety of $L_X$ at fixed mass that is observed in XUCS,
they offer a spread that is large enough to study the origin of the spread and to characterise gas-poor haloes. 

\begin{figure}
\centerline{\includegraphics[width=1.\linewidth]{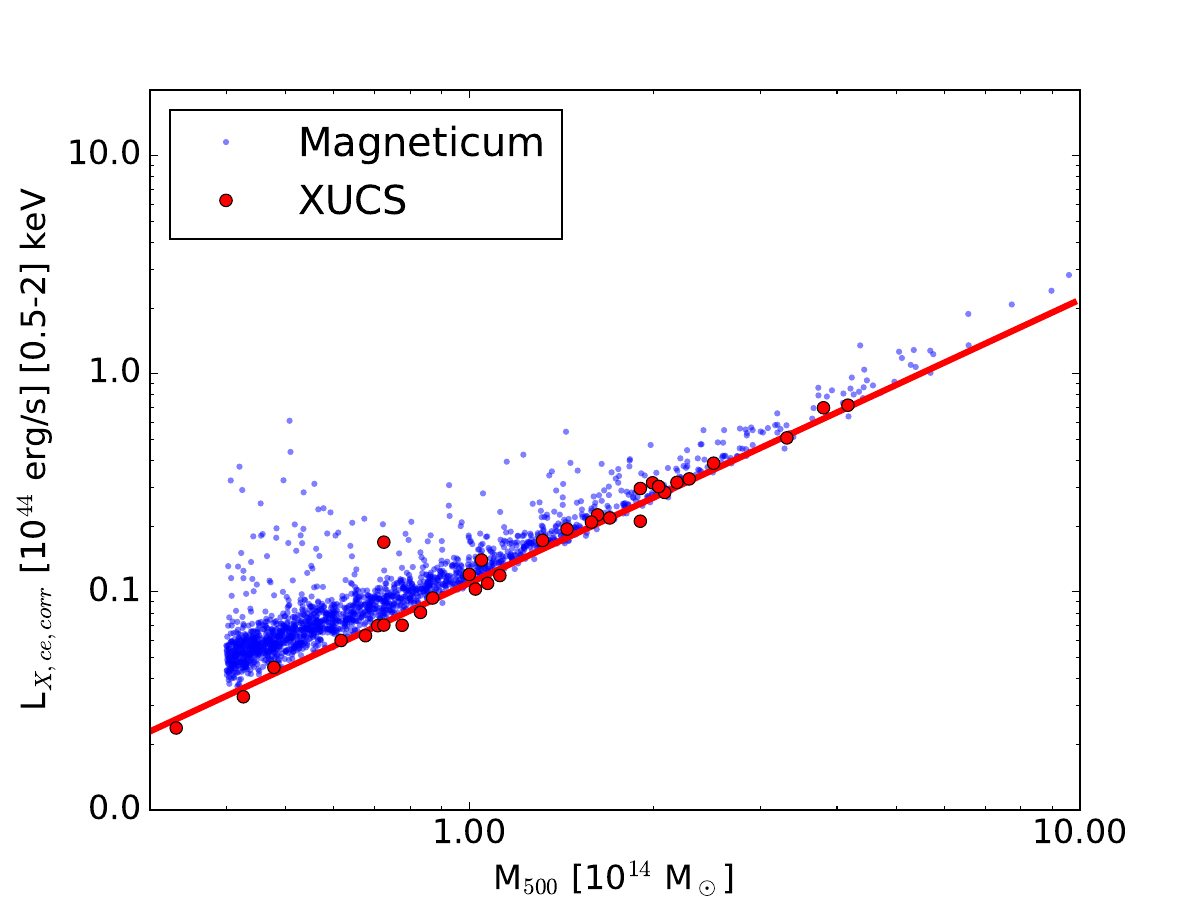}}
\caption[h]{Core-excised X-ray luminosity corrected for gas fraction $L_{X, \rm ce,corr}$  (see Eq. \ref{eq:lxcorr}) $vs.$ mass $M$. The gas-fraction
correction removes most of the scatter around the relation (compare with
Fig. \ref{fig:LxM}). }
\label{fig:LxMgcolorcor}
\end{figure}

Figure \ref{fig:fgasM} shows the $f_g-M$ relation for the simulated data for XUCS and for an X-ray selected sample of relaxed clusters \citep{2006ApJ...640..691Vikhlinin,2009ApJ...693.1142Sun}. 
Simulations and real data share similar slopes (Magneticum: 0.21, and V+S: 0.15; \citealt{2010MNRAS.407..263Andreon}). The XUCS slope is inherited from that of V+S; see \cite{2017A&A...606A..24Andreon}. 
The scatter at fixed mass of the simulated clusters
is 0.08 dex (at the high-mass end) to 0.11 dex (in the low-mass range), intermediate between the XUCS intrinsic scatter (0.17 dex), but larger than that of V+S  \citep[0.06 dex;][]{2010MNRAS.407..263Andreon}. %(0.06 dex, A10 ref). 
The V+S sample has the largest intercept and lowest  scatter, which are two known features of samples selected in this way~\citep{2017A&A...606A..24Andreon}.
The Magneticum mean relation has a lower intercept than the X-ray selected sample, 
but it is higher than XUCS, indicating that clusters are more gas poor in the Universe than in Magneticum.
Although the spread at a fixed mass of the simulated data  for $L_X$ does not exactly match that of XUCS, it is large enough for us to investigate the origin of the spread. 

\subsection{Dependence on gas fraction, richness, concentration, and formation redshift}
\label{sec:gfc}

\begin{figure*}
\centerline{
\includegraphics[width=.33\linewidth]{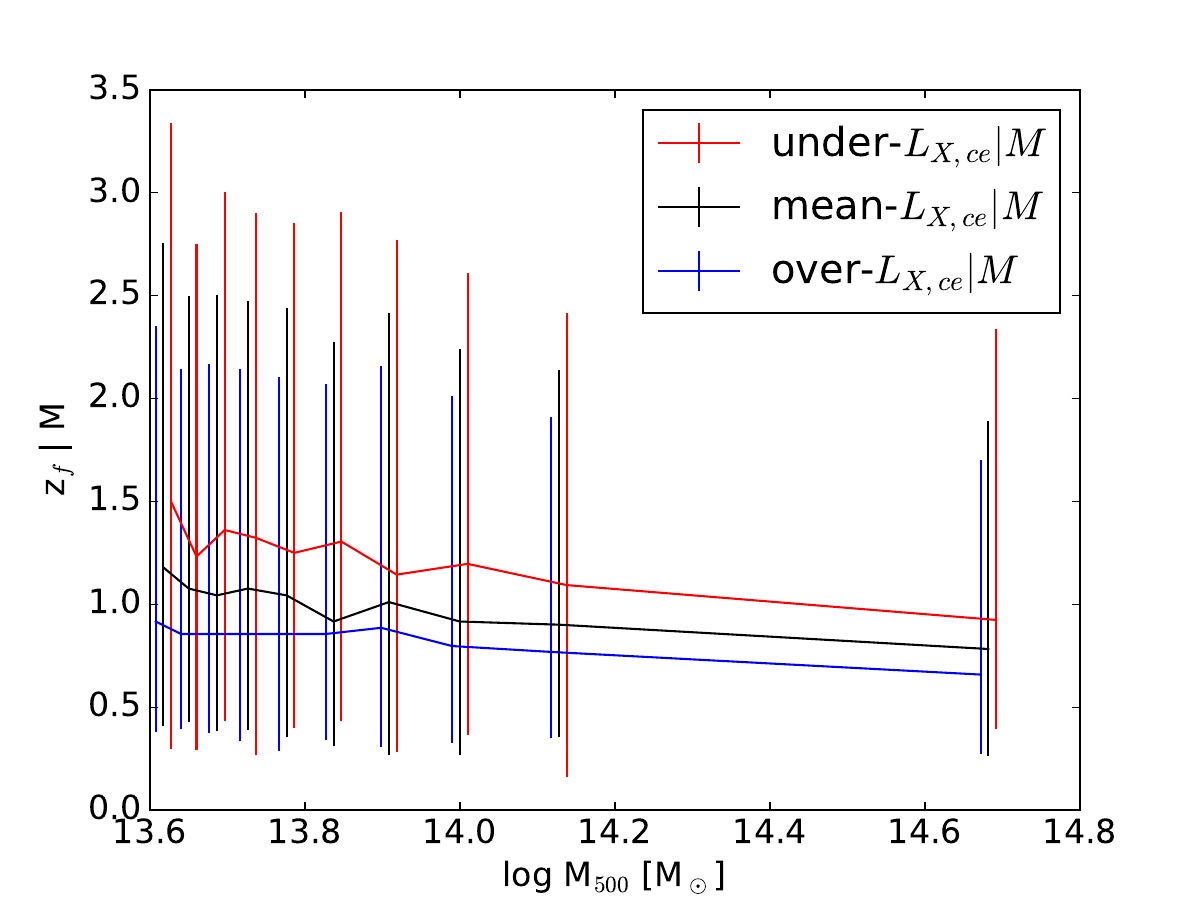}
\includegraphics[width=.33\linewidth]{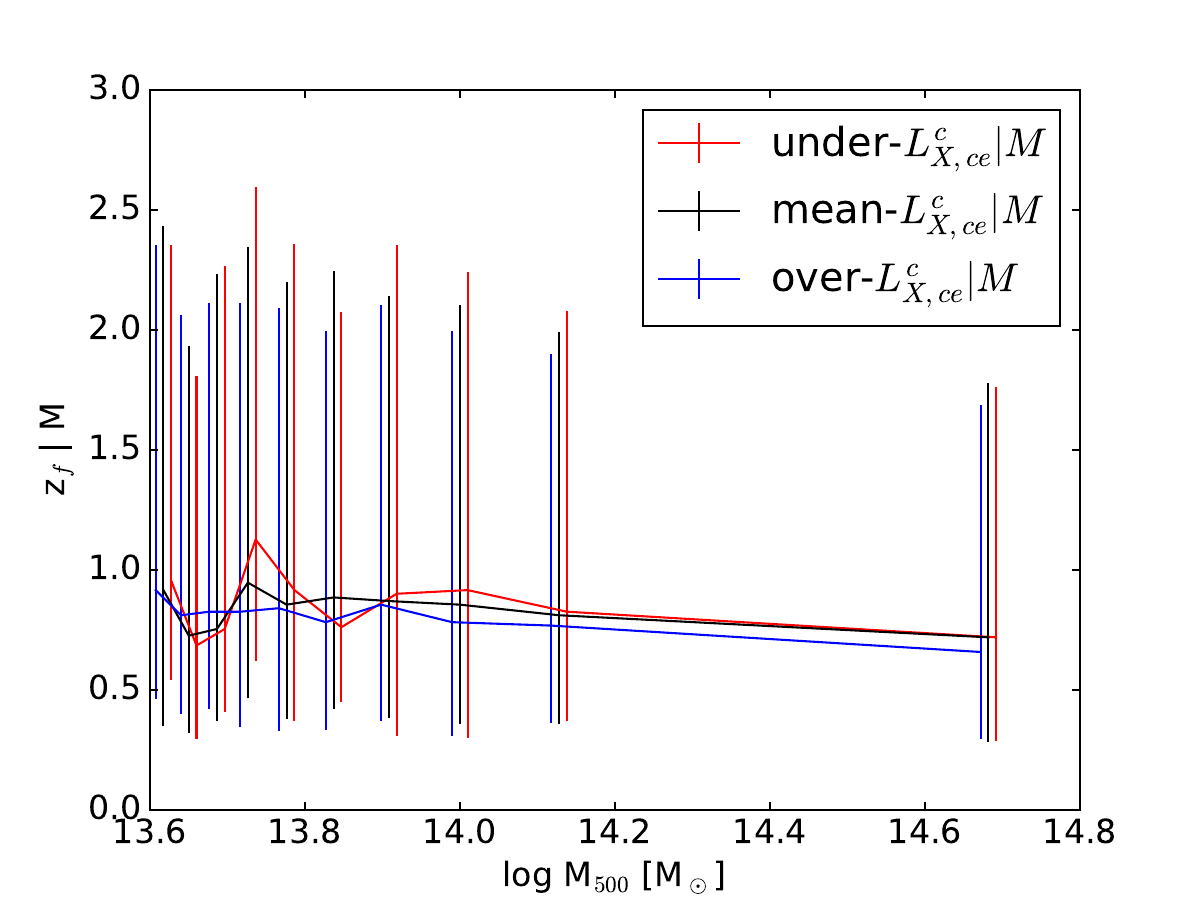}
\includegraphics[width=.33\linewidth]{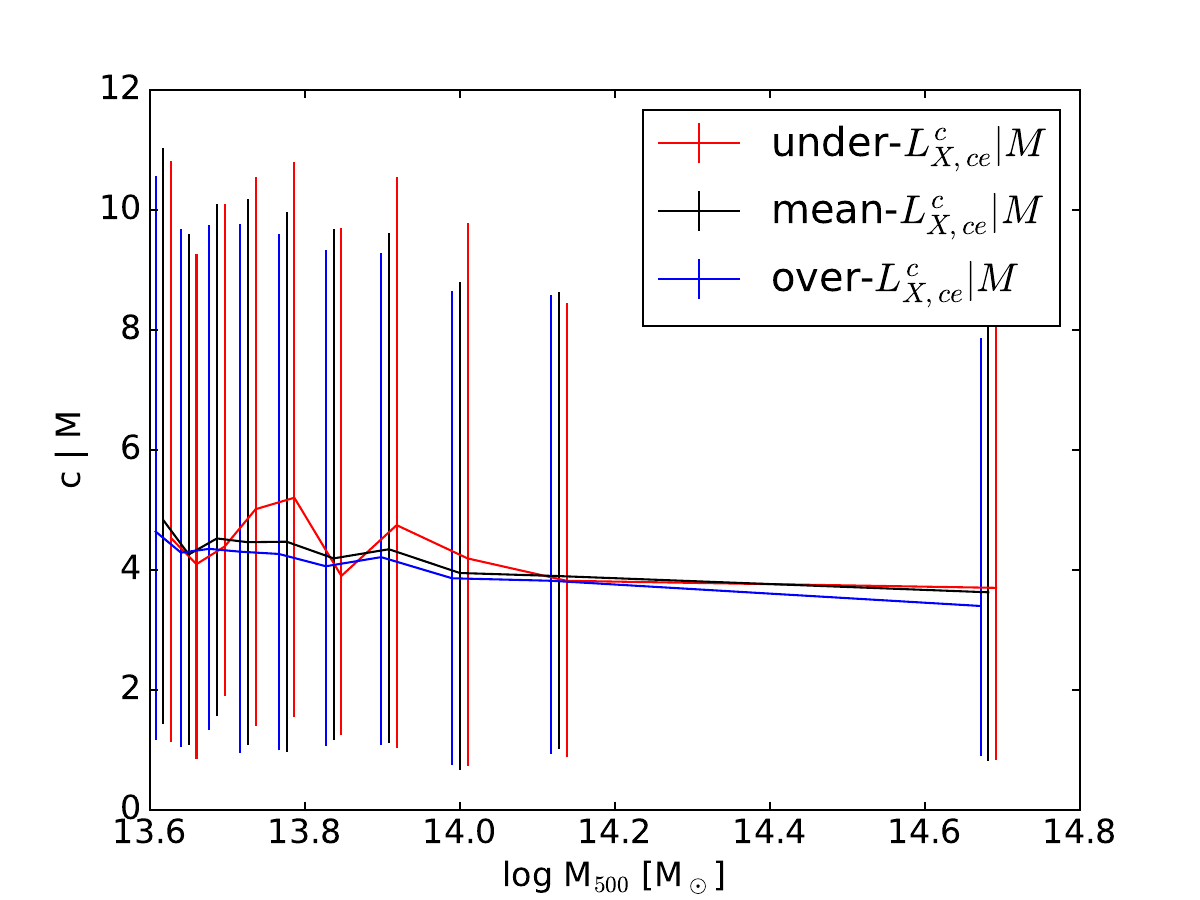}
}
\caption[h]{Covariance between formation redshift and X-ray luminosity residuals, with and without gas fraction corrections. Left\LEt{please add a short descriptive sentence omitting the initial article before you describe the individual panels. This applies throughout, please check and change as required} panel: Formation redshift (16,50,86 percentile) at fixed mass for clusters with $L_{X,ce}$ that is not corrected for gas, split into three tertiles. Middle panel: Formation redshift (16,50,86 percentile) at fixed mass for a cluster with gas-corrected $L_{X,ce}$ split into three tertiles.
 Right panel: Concentration (16,50,86 percentile) at fixed mass for a cluster with gas-corrected $L_{X,ce}$ split into three tertiles.
 }
\label{fig:zf}
\end{figure*}

Figure \ref{fig:LxMcolor} shows $L_{X, \rm{ce}}$ $\text{versus}$ $M$ of Magneticum galaxy clusters colour-coded by $f_g$ (first panel), formation redshift (second panel), concentration (third panel), richness (fourth panel), and BCG stellar mass (fifth panel). The top panel shows that a fixed mass, brighter clusters are gas rich,
in agreement with the observational result in \cite{2017A&A...606A..24Andreon}. They were presented in a similar plot in
\cite{2022MNRAS.511.4991Andreon}.
The other panels illustrate that at fixed mass, the X-ray luminosity is correlated to all plotted quantities
(blue points are systematically above or below red points), as quantified in Sec.~\ref{sec:cov-full} : 
LSB clusters are older (second panel from the top), concentrated 
(central panel), are less rich (second panel from the bottom),  and have more massive  BCGs (bottom panel).

\cite{2017A&A...606A..24Andreon}  showed that the $L_X-M$ scatter can be drastically reduced when the X-ray luminosities are corrected for differences in gas fraction; see the red points in Fig.~\ref{fig:LxMgcolorcor} 
We now determine whether this is true for the simulations as well.
For this reason, we defined for each galaxy cluster the  $L_{X, \rm{ce}}$ luminosity corrected ($L_{X, \rm{ce, corr}}$) for its $f_g$ as \begin{equation}
\label{eq:lxcorr}
L_{X, \rm{ce, corr}} = 
    \left(
        \dfrac{f_g\left(M\right)}{f_g} 
    \right)^2 L_{X, \rm{ce}},
\end{equation}
where $f_g\left(M\right)$ is the average gas fraction at its mass $M,$ estimated via a linear regression of $\log f_g$ $\text{versus}$ $\log M$. Fig. \ref{fig:LxMgcolorcor} shows that
the scatter of the luminosity--mass relation 
is drastically reduced
compared to Fig. \ref{fig:LxM}. The scatter decreases from a range of $0.24-0.30$ dex (before correction) to $0.09-0.11$ dex (after correction).
It is now readily apparent that the Magneticum and XUCS X-ray $L_X-M$ scaling relations are quite similar, but have  a slightly different   slope and scatter. Nevertheless, both scaling relations become much tighter after the gas fraction is corrected for.

The question is whether all the covariances illustrated in Fig.~\ref{fig:LxMcolor} disappear when gas-fraction corrected
X-ray luminosities are employed. Sec.~\ref{sec:cov-full} quantitatively addresses this point, but we
illustrate the effect of the gas-fraction correction on the concentration and formation redshift here. When the clusters are split into tertilies of X-ray luminosities at fixed mass, LSB clusters have older formation times (left panel of  
figure \ref{fig:zf}). However,  when we use  tertiles of X-ray corrected luminosities ($L_{X, \rm{ce, corr}}$),  then the formation times (central panel) and concentration (right panel)   show no covariance with $L_{X, \rm{ce, corr}}.$

\subsection{Full covariance analysis}
\label{sec:cov-full}

\begin{figure*}
\centerline{
\includegraphics[width=0.9\linewidth]{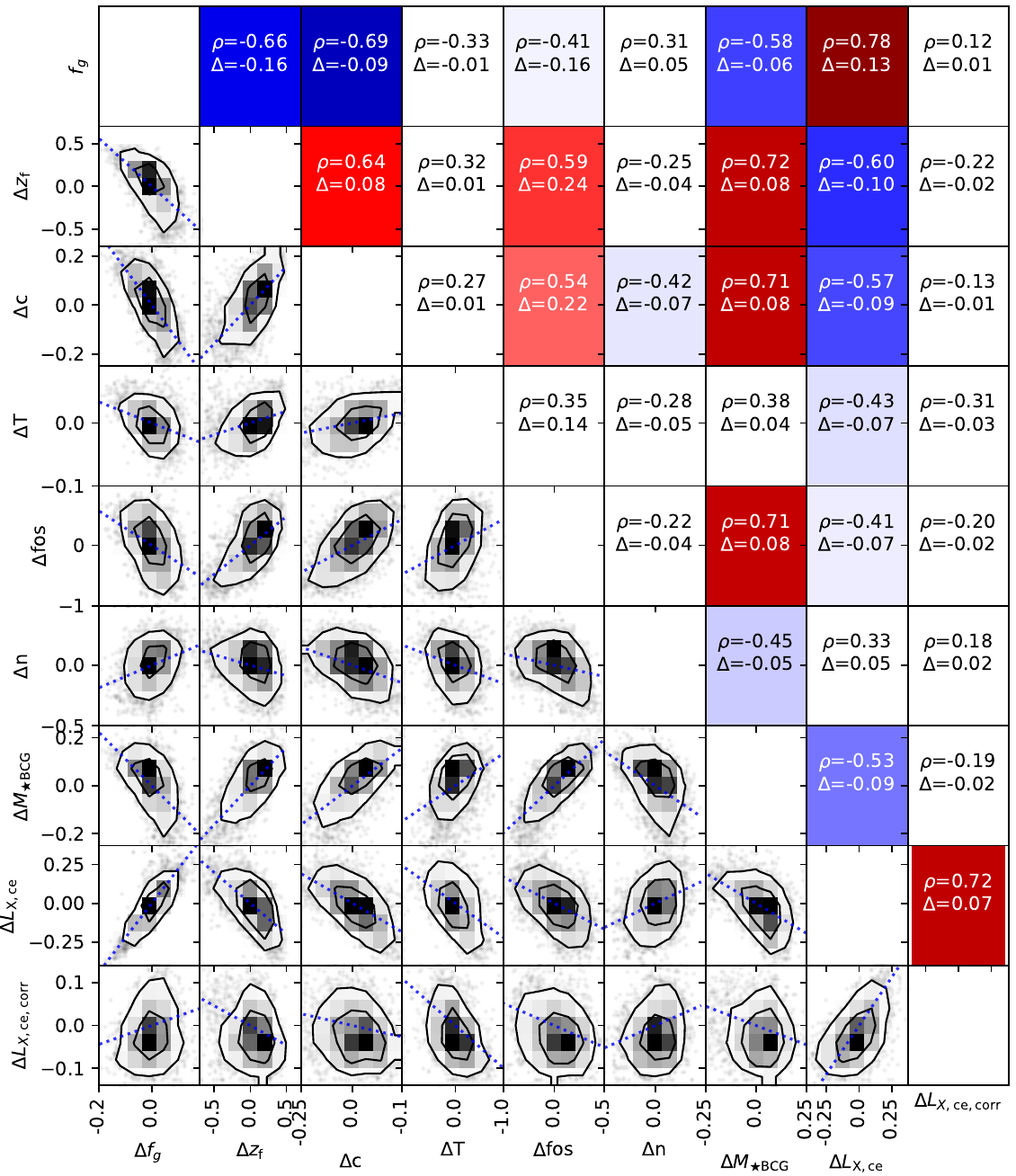}
}
\caption[h]{Correlation between cluster properties. Lower triangle: Scatter plot of the residuals at fixed halo mass $M_{500c}$ of $\log_{10}$ gas fraction $f_g,$ $\log_{10}$ formation redshift $z_f,$ $\log_{10}$ concentration $c$, $\log_{10}$ temperature, $\log_{10}$ fossilness, $\log_{10}$ richness $n,$  $\log_{10}$ X-ray luminosity, and $\log_{10}$ $f_g^2$ corrected luminosity (as in Eq. \ref{eq:lxcorr}). The dotted lines show the slope  for individual correlation coefficients. When it points in a direction different from the data elongation, it indicates that the data do not satisfy the assumption for the derivation of the correlation coefficient.
The upper triangle shows the corresponding correlation coefficient $\rho$ and the amplitude of the covariance $\Delta$. 
}
\label{fig:covar-scat}
\end{figure*}

To study the covariance at fixed mass, we
performed a linear regression of the logarithm of each property ($f_g, z_f, c, T,$ fossilness,  $n, M_{\star \rm BCG}, L_{\rm X,ce},$ and $L_{\rm X,ce,corr}$) against $\log $M, and we searched for covariance in the residuals from the mean
relation.
Fig.~\ref{fig:covar-scat} shows the correlation coefficient $\rho$ (upper triangle) and residual scatter (lower triangle) plots. The first is a measure of the linear correlation between residuals, and gives the fraction of the
ordinate scatter explained by the abscissa scatter given the linear slope $s$ for  individual correlation coefficients, $s=\rho \sigma_y/\sigma_x.$ 
 This is shown as the dashed line in the lower triangle plots.
The lower triangle plots illustrate whether the linearity hypothesis is plausible, and also whether the trend is of any physical significance: a correlation involving a negligible change is unlikely to be of any physical interest, even if it is statistically significant. We quantify the above with the ordinate variation when the abscissa changes by $1\sigma$, $\Delta = s \sigma_x = \rho \sigma_y$, whose values are reported in the upper triangle. 

Values of $\Delta<0.05$ (i.e. $\sim10$\% differences) are hardly accessible with observations of individual objects in the large samples needed for covariance studies, whereas correlations $\rho<0.3$ often result from assumptions that are not satisfied by the data. The richness-temperature covariance for example is vertically elongated, but the correlation
coefficient indicates an elongation of 45 degree away from it (see Fig~\ref{fig:covar-scat}). These correlation coefficients should be considered with caution.

Fig.~\ref{fig:covar-scat} confirms the tight correlation at fixed mass of the core-excised X-ray luminosity
with gas-fraction (i.e. that gas-poor clusters are also X-ray faint; this covariance has the highest correlation coefficient, 0.78), with formation redshift (i.e, that
these X-ray faint clusters formed early) and with concentration (i.e. that X-ray faint clusters have a high concentration). 

The values of $L_{X, \rm{ce, corr}}$ show little or no covariance with any other property ($\rho\lesssim 0.3$), as already found in the previous section for some of these quantities. Thus the gas-fraction correction absorbs most of the scatter in all the considered scaling relations.

This is remarkable because this correction was motivated by removing covariance with one quantity, gas fraction, not with the remaining six quantities. The disappearance of these correlations indicates a common origin for the residuals from the mass trend, that is, that the physical reason for the differences in gas fraction also likely causes deviations from the other quantities from the mean relation.

Out of the remaining quantities, three of the largest covariances are with BGC stellar mass ($vs.$ with formation redshift, concentration, and fossilness) in the sense that clusters of a given mass with more massive
BCGs also have a higher-than-average formation redshift and a higher-than-average concentration. We interpret this result as due to the longer time that the BCG had to assimilate more satellites~\citep{2019MNRAS.486.4001Ragagnin}.  
Fossilness, concentration, and formation time show strong covariance~\citep{2019MNRAS.486.4001Ragagnin}. Finally, gas fraction and formation times are anti-correlated, which may impact observational studies that constrain cosmological parameters with gas fraction~\citep[e.g.][]{2003A&A...398..879Ettori,2022MNRAS.510..131Mantz}. 
Interestingly, richness shows low levels of covariance ($\Delta$) with formation redshift, gas fraction, and a weak correlation with  concentration and mass of the BCG. This confirms that it can be a very useful mass proxy with little sensitivity to the mass accretion history, differently from, for instance,  gas fraction or X-ray luminosity.
Of the considered quantities, the temperature ist less covariant with the others: all correlation values are
below  $\approx0.4,$  and some of the covariances (with gas fraction, concentration, and gas fraction) are hardly observable at best.
The source of the temperature variance therefore
appears to be unrelated to the others we considered.

\section{Discussion}
\label{sec:discu}

\subsection{Comparison with observations and previous simulations}
\label{sec:compa}

Our analysis extends previous simulation works that investigated one or just a few covariance quantities \citep{2002ApJ...568...52Wechsler,2003MNRAS.339...12Zhao,2006MNRAS.368.1931Lu,2010MNRAS.407..581RagoneFigueroa,2011MNRAS.416.2997Cui,2012MNRAS.426.2046Angulo,2012MNRAS.422..185Giocoli,2019MNRAS.486.4001Ragagnin,2019MNRAS.490.5693Bose,2020MNRAS.495..686Anbajagane,2020MNRAS.498.4450Wang,2021arXiv211204926Richardson} by addressing a larger set of variables that is analysed with a single numerical technique.
Moreover, we tested the finding of recent observational studies~\citep{2017A&A...606A..24Andreon}   that most of the scatter of the $L_x-M$ relation can be absorbed by correcting $L_x$ for $f_g^2.$

In observational studies, \cite{2019NatCo..10.2504Farahi} (that focuses on the most massive clusters), found a negative correlation between hot ($f_g$) and cold (richness) baryons at fixed mass, unlike \cite{2022MNRAS.511.2968Puddu}, who used XUCS clusters. 
Our simulated data agree with \cite{2022MNRAS.511.2968Puddu} that gas fraction and richness show a positive  correlation.  
However,  the way in which \cite{2022MNRAS.511.2968Puddu} computed richnesses differs from our way. The authors counted galaxies within\LEt{may I suggest that you run a spellchecker before you submit your next paper} a cylinder, not a sphere, and used a  slightly higher galaxy mass threshold.

On the other hand, \cite{2022MNRAS.511.2968Puddu} reported that LSB clusters have fainter BCGs and a slightly smaller magnitude gap ($\Delta M_{1,2}$ between the central and the brightest satellite), while the Magneticum simulation shows an opposite trend according to which gas-poor clusters are old (see Fig. \ref{fig:covar-scat}) 
and have  brighter BCGs, because the BCG had time to accrete more satellites~\citep[as in Fig. 7 and Fig. 8 in][]{2019MNRAS.486.4001Ragagnin}. We emphasise, however, that separating the BCG luminosity from intracluster light is challenging and that different methods have been used to address this problem in simulations and observations. 

Gas-poor clusters have a high concentration in simulations on average (with some
scatter, see Fig.~\ref{fig:covar-scat}).
CL2015 \citep{2019A&A...630A..78Andreon} is a gas-poor galaxy cluster with a low concentration. Although it does not follow the average trend seen in simulation, it lies within the $1\sigma$ scatter of it.

\begin{figure}
\centerline{
\includegraphics[width=0.9\linewidth]{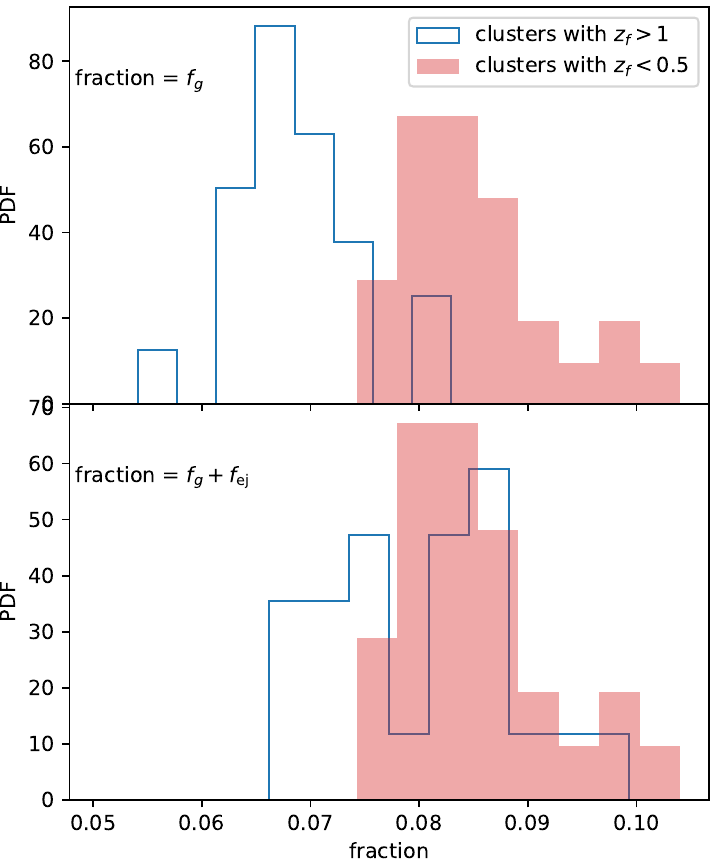}
}
\caption[h]{PDF of gas fraction for old clusters ($z_f>1$, with blue steps) and young clusters ($z_f<0.5$, as pink shaded area). The top panel shows the gas-fraction within $R_{\rm 500c}$ and the bottom panel shows the gas-fraction within $R_{\rm 500c}$ corrected with the amount of gas  that was inside the halo at formation time and have been ejected outside outside the virial radius  ($f_{\rm ej}$).
 In both cases we consider haloes in the range of mass $M_{\rm 200c}\in[1-2]\times 10^{14}M_\odot.$
 We see that the amount of ejected gas accounts for much of the difference in $f_g$ between old and young clusters.}
\label{fig:stellarmass}
\end{figure}

\subsection{Role of AGN outflow}

Full-physics simulations and   processes such as AGN feedback proved to significantly impact galaxy cluster   mass profiles  up to their outskirts ~\citep{2010MNRAS.405.2161Duffy,2011MNRAS.416..801Fabjan,2014MNRAS.442.2641Velliscig}, to lower the gas fraction, to   strongly suppress star formation  galaxies~\citep{2017MNRAS.465...32Bower}, and to boost the  X-ray luminosity during mergers~\citep[see e.g.][]{2004MNRAS.349..476Torri,2007MNRAS.380..437Poole}.

Feedback from AGN in particular is capable of ejecting a significant fraction of gas outside the halo~\citep{2020MNRAS.491.4462Davies}. We studied whether AGN feedback is the reason that older clusters are gas poor.
To this end, we selected clusters in the range of $M_{\rm 200c}\in[1-2]\times10^{14}{\rm M}_\odot$ 
and followed  gas particles in time that were inside the virial radius at formation time.

Fig. \ref{fig:stellarmass} (top panel) shows that the gas fraction 
     of old clusters ($z_f>1,$ blue histogram) is lower than for young clusters ($z_f<0.5,$ pink shaded histogram). The difference is mainly due to gas that was within the virial radius at formation time and that  was later ejected ($f_{\rm ej}$). When the ejected gas is returned, the two distributions approach each other
(bottom panel of Fig. \ref{fig:stellarmass})

The increase in $f_g$ that we find between the top and bottom panels of Fig.~\ref{fig:stellarmass} is not an artefact because   $\rho_c$ changes with time (which enters in the definition of $R_{\rm 200c}$).
 $\rho_c$ decreases with time, and the baryon fraction of the material  that is accreted by clusters is similar to the cosmic baryon fraction~\citep[see Sec. 4.3  in][]{2020MNRAS.499.2303VallesPerez}.
Moreover, the conversion of gas into stars plays 
a negligible role, as we estimated that the fraction of stars that is produced after formation time is $<5\%$ of  the galaxy cluster final masses at $z=0$.

\begin{figure*}
\centerline{
\includegraphics[width=\linewidth]{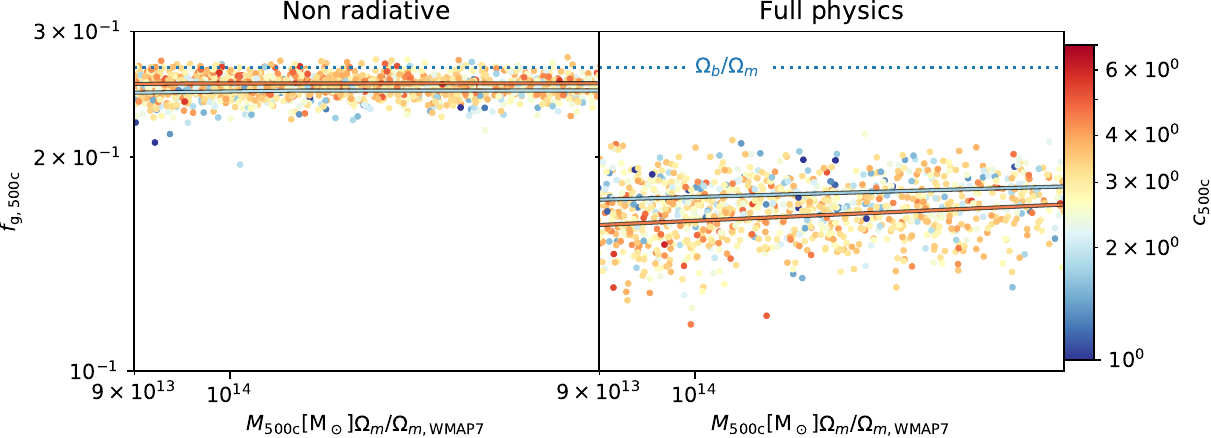}
}
\caption[h]{Gas fraction $f_g$ $vs.$ halo mass $M_{\rm 500c}$ for Magneticum Box1a/mr non-radiative run (left panel) and full-physics run (right panel). Points are colour-coded by concentration within $R_{\rm 500c}.$  Lines show the average gas fraction for the two quantiles in gas concentration, up to $16\%$ in blue, and from $85\%$ in red, both coloured according to their average concentration. The non-radiative simulation shows a slight dependence of concentration on gas fraction, while full-physics runs show the opposite behaviour: the gas fraction is anti-correlated with concentration. Masses are corrected to WMAP7 $\Omega_m$ for consistency with the other results of this study.
}
\label{fig:fgnorad}
\end{figure*}

\subsection{Role of baryon physics in the  gas fraction $\text{versus}$ concentration covariance}

In this subsection we rule out the possibility that the correlation between gas fraction and formation redshift is due to environmental effects in older galaxy clusters.
To this end, we studied the relation between gas fraction and concentration (which correlates with formation redshift) at fixed mass by comparing  a non-radiative run  with  a full-physics run.
We do not have a WMAP7 non-radiative counterpart; therefore we compared two Magneticum Box1a/mr\footnote{
Simulations are presented in \cite{2020MNRAS.494.3728Singh}, cover a volume of $\approx1300$ comoving Mpc,  have a dark matter particle mass of $m_{\rm DM}=2\times 10^{10} M_\odot,$ a gas mass $m_{\rm DM}=4\times 10^{9} M_\odot,$ and a softening $\epsilon=15.5$ comoving kpc for both dark matter and gas. }
simulations with cosmological parameters   $\Omega_m=0.153,\Omega_b=0.0408,\sigma_8=0.614,$ and $h_0=0.666.$

Figure \ref{fig:fgnorad} shows $f_g$ as a function of halo mass colour-coded by halo concentration taken from \cite{2021MNRAS.500.5056Ragagnin}. In the non-radiative run, the gas-fraction is close to the cosmic gas fraction $\Omega_b/\Omega_m,$  while full-physics simulations have much lower $f_g$, in agreement with  previous studies~\citep{2011A&A...526A..79Eckert,2013MNRAS.431.1487Planelles}, which showed that this difference is due to the AGN feedback, because the ejected gas is capable of escaping  from  clusters.
The non-radiative simulation  $f_g-M$ relation has an indication of an anti-correlation with concentration at most, that is, gas-rich clusters tend to be concentrated (left panel), as opposed to the full-physics run (right panel).
We can speculate that in non-radiative simulations,  due to the lack of AGN feedback, the process of adiabatic contraction brings more gas to the center as the dark matter halo becomes more concentrated.
This experiment shows that very different baryon-physics prescriptions can reverse the correlation between $c$ and $f_g,$ which explains why the~\cite{2010ApJ...715.1508Stanek} simulations and modern simulations reported contrasting results.

 \section{Conclusion}
\label{sec:conclu}

We characterised the origin of LSB and gas-poor galaxy clusters in cosmological simulations by analysing several haloes from the Magneticum   simulations. \LEt{please remove the bullet points below so that this reads as proper paragraphs}. We found a strong covariance between the X-ray luminosity and the gas fraction at fixed mass, in agreement with recent studies of unbiased samples (XUCS) from \cite{2022MNRAS.511.2968Puddu}, and following the reasoning in \cite{2017A&A...606A..24Andreon} that the  X-ray luminosity scales with $f_g^2$ at fixed halo mass,
we were able to remove the X-ray intrinsic scatter by applying a gas-fraction correction in both simulations and observations (see our Fig. \ref{fig:LxMgcolorcor}).
In particular, correcting the luminosity for  $f_g^2$ reduced all our correlation coefficients to almost zero (see last column in Fig. \ref{fig:covar-scat}).
  
  We then characterised  LSB and gas-poor  galaxy clusters and found them  to be older, with a higher concentration, and with a slight anti-correlation with richness and tested whether the feedback mechanism lowers the gas fraction of older clusters.
To this end, we quantified the amount of depleted gas at formation time and found that  older clusters depleted more gas and that the amount of depleted gas is enough to justify their low gas fraction (compared to gas-rich systems in a fixed mass bin).
 Finally, we ruled out the hypothesis that the environment of older clusters caused them to be gas poor. To do this, we compared full-physics simulations against non-radiative simulations, and  we found that  the latter do not show any negative correlation between gas fraction and concentration (which strongly correlates with formation redshift).
We therefore conclude that older galaxy clusters tend to be gas poor (and thus LSB) because the feedback mechanism depleted a significant amount of gas from these systems.

\section*{Acknowledgements}
The \textit{Magneticum Pathfinder} simulations were partially performed at the Leibniz-Rechenzentrum with CPU time assigned to the Project `pr86re'.   AR acknowledges support from the grant PRIN-MIUR 2017 WSCC32 and acknowledges the usage of the INAF-OATs IT framework~\citep{2020ASPC..527..307Taffoni,2020ASPC..527..303Bertocco}.

\section*{Data availability}
Raw simulation data were generated at the C$^2$PAP/LRZ cosmology simulation web portal \url{https://c2papcosmosim.uc.lrz.de/}. The derived data supporting the findings of this study are available from the corresponding author AR on request. 
The derived data supporting the findings of this study are available from the corresponding author SA on request.

\bibliographystyle{aa} % style aa.bst
 \bibliography{referenze} % your references Yourfile.bib

\begin{thebibliography}{102}
\expandafter\ifx\csname natexlab\endcsname\relax\def\natexlab#1{#1}\fi

\bibitem[{{Allen} {et~al.}(2011){Allen}, {Evrard}, \&
  {Mantz}}]{2011ARA&A..49..409Allen}
{Allen}, S.~W., {Evrard}, A.~E., \& {Mantz}, A.~B. 2011, \araa, 49, 409

\bibitem[{{Anbajagane} {et~al.}(2020){Anbajagane}, {Evrard}, {Farahi},
  {Barnes}, {Dolag}, {McCarthy}, {Nelson}, \&
  {Pillepich}}]{2020MNRAS.495..686Anbajagane}
{Anbajagane}, D., {Evrard}, A.~E., {Farahi}, A., {et~al.} 2020, \mnras, 495,
  686

\bibitem[{{Andreon}(2010)}]{2010MNRAS.407..263Andreon}
{Andreon}, S. 2010, \mnras, 407, 263

\bibitem[{{Andreon} {et~al.}(2016){Andreon}, {Dong}, \&
  {Raichoor}}]{2016A&A...593A...2Andreon}
{Andreon}, S., {Dong}, H., \& {Raichoor}, A. 2016, \aap, 593, A2

\bibitem[{{Andreon} \& {Moretti}(2011)}]{2011A&A...536A..37Andreon}
{Andreon}, S. \& {Moretti}, A. 2011, \aap, 536, A37

\bibitem[{{Andreon} {et~al.}(2019){Andreon}, {Moretti}, {Trinchieri}, \&
  {Ishwara-Chandra}}]{2019A&A...630A..78Andreon}
{Andreon}, S., {Moretti}, A., {Trinchieri}, G., \& {Ishwara-Chandra}, C.~H.
  2019, \aap, 630, A78

\bibitem[{{Andreon} {et~al.}(2022){Andreon}, {Trinchieri}, \&
  {Moretti}}]{2022MNRAS.511.4991Andreon}
{Andreon}, S., {Trinchieri}, G., \& {Moretti}, A. 2022, \mnras, 511, 4991

\bibitem[{{Andreon} {et~al.}(2017{\natexlab{a}}){Andreon}, {Trinchieri},
  {Moretti}, \& {Wang}}]{2017A&A...606A..25Andreon}
{Andreon}, S., {Trinchieri}, G., {Moretti}, A., \& {Wang}, J.
  2017{\natexlab{a}}, \aap, 606, A25

\bibitem[{{Andreon} {et~al.}(2017{\natexlab{b}}){Andreon}, {Wang},
  {Trinchieri}, {Moretti}, \& {Serra}}]{2017A&A...606A..24Andreon}
{Andreon}, S., {Wang}, J., {Trinchieri}, G., {Moretti}, A., \& {Serra}, A.~L.
  2017{\natexlab{b}}, \aap, 606, A24

\bibitem[{{Angulo} {et~al.}(2012){Angulo}, {Springel}, {White}, {Jenkins},
  {Baugh}, \& {Frenk}}]{2012MNRAS.426.2046Angulo}
{Angulo}, R.~E., {Springel}, V., {White}, S.~D.~M., {et~al.} 2012, \mnras, 426,
  2046

\bibitem[{{Arnaud}(1996)}]{1996ASPC..101...17Arnaud}
{Arnaud}, K.~A. 1996, in Astronomical Society of the Pacific Conference Series,
  Vol. 101, Astronomical Data Analysis Software and Systems V, ed. G.~H.
  {Jacoby} \& J.~{Barnes}, 17

\bibitem[{{Arnaud} {et~al.}(2010){Arnaud}, {Pratt}, {Piffaretti},
  {B{\"o}hringer}, {Croston}, \& {Pointecouteau}}]{2010A&A...517A..92Arnaud}
{Arnaud}, M., {Pratt}, G.~W., {Piffaretti}, R., {et~al.} 2010, \aap, 517, A92

\bibitem[{{Beck} {et~al.}(2016){Beck}, {Murante}, {Arth}, {Remus}, {Teklu},
  {Donnert}, {Planelles}, {Beck}, {F{\"o}rster}, {Imgrund}, {Dolag}, \&
  {Borgani}}]{2016MNRAS.455.2110Beck}
{Beck}, A.~M., {Murante}, G., {Arth}, A., {et~al.} 2016, \mnras, 455, 2110

\bibitem[{{Beltz-Mohrmann} \&
  {Berlind}(2021)}]{2021ApJ...921..112BeltzMohrmann}
{Beltz-Mohrmann}, G.~D. \& {Berlind}, A.~A. 2021, \apj, 921, 112

\bibitem[{{Bertocco} {et~al.}(2020){Bertocco}, {Goz}, {Tornatore}, {Ragagnin},
  {Maggio}, {Gasparo}, {Vuerli}, {Taffoni}, \&
  {Molinaro}}]{2020ASPC..527..303Bertocco}
{Bertocco}, S., {Goz}, D., {Tornatore}, L., {et~al.} 2020, in Astronomical
  Society of the Pacific Conference Series, Vol. 527, Astronomical Society of
  the Pacific Conference Series, ed. R.~{Pizzo}, E.~R. {Deul}, J.~D. {Mol},
  J.~{de Plaa}, \& H.~{Verkouter}, 303

\bibitem[{{Biffi} {et~al.}(2013){Biffi}, {Dolag}, \&
  {B{\"o}hringer}}]{2013MNRAS.428.1395Biffi}
{Biffi}, V., {Dolag}, K., \& {B{\"o}hringer}, H. 2013, \mnras, 428, 1395

\bibitem[{{Bocquet} {et~al.}(2016){Bocquet}, {Saro}, {Dolag}, \&
  {Mohr}}]{2016MNRAS.456.2361Bocquet}
{Bocquet}, S., {Saro}, A., {Dolag}, K., \& {Mohr}, J.~J. 2016, \mnras, 456,
  2361

\bibitem[{{Bode} {et~al.}(2009){Bode}, {Ostriker}, \&
  {Vikhlinin}}]{2009ApJ...700..989Bode}
{Bode}, P., {Ostriker}, J.~P., \& {Vikhlinin}, A. 2009, \apj, 700, 989

\bibitem[{{B{\"o}hringer} {et~al.}(2007){B{\"o}hringer}, {Schuecker}, {Pratt},
  {Arnaud}, {Ponman}, {Croston}, {Borgani}, {Bower}, {Briel}, {Collins},
  {Donahue}, {Forman}, {Finoguenov}, {Geller}, {Guzzo}, {Henry}, {Kneissl},
  {Mohr}, {Matsushita}, {Mullis}, {Ohashi}, {Pedersen}, {Pierini}, {Quintana},
  {Raychaudhury}, {Reiprich}, {Romer}, {Rosati}, {Sabirli}, {Temple}, {Viana},
  {Vikhlinin}, {Voit}, \& {Zhang}}]{2007A&A...469..363Boehringer}
{B{\"o}hringer}, H., {Schuecker}, P., {Pratt}, G.~W., {et~al.} 2007, \aap, 469,
  363

\bibitem[{{Bose} {et~al.}(2019){Bose}, {Eisenstein}, {Hernquist}, {Pillepich},
  {Nelson}, {Marinacci}, {Springel}, \&
  {Vogelsberger}}]{2019MNRAS.490.5693Bose}
{Bose}, S., {Eisenstein}, D.~J., {Hernquist}, L., {et~al.} 2019, \mnras, 490,
  5693

\bibitem[{{Bower} {et~al.}(2017){Bower}, {Schaye}, {Frenk}, {Theuns},
  {Schaller}, {Crain}, \& {McAlpine}}]{2017MNRAS.465...32Bower}
{Bower}, R.~G., {Schaye}, J., {Frenk}, C.~S., {et~al.} 2017, \mnras, 465, 32

\bibitem[{{Boylan-Kolchin} {et~al.}(2009){Boylan-Kolchin}, {Springel}, {White},
  {Jenkins}, \& {Lemson}}]{2009MNRAS.398.1150Boylan}
{Boylan-Kolchin}, M., {Springel}, V., {White}, S. D.~M., {Jenkins}, A., \&
  {Lemson}, G. 2009, \mnras, 398, 1150

\bibitem[{{Castro} {et~al.}(2021){Castro}, {Borgani}, {Dolag}, {Marra},
  {Quartin}, {Saro}, \& {Sefusatti}}]{2021MNRAS.500.2316Castro}
{Castro}, T., {Borgani}, S., {Dolag}, K., {et~al.} 2021, \mnras, 500, 2316

\bibitem[{{Corasaniti} {et~al.}(2021){Corasaniti}, {Sereno}, \&
  {Ettori}}]{2021ApJ...911...82Corasaniti}
{Corasaniti}, P.-S., {Sereno}, M., \& {Ettori}, S. 2021, \apj, 911, 82

\bibitem[{{Crain} {et~al.}(2007){Crain}, {Eke}, {Frenk}, {Jenkins}, {McCarthy},
  {Navarro}, \& {Pearce}}]{2007MNRAS.377...41Crain}
{Crain}, R.~A., {Eke}, V.~R., {Frenk}, C.~S., {et~al.} 2007, \mnras, 377, 41

\bibitem[{{Cui} {et~al.}(2011){Cui}, {Springel}, {Yang}, {De Lucia}, \&
  {Borgani}}]{2011MNRAS.416.2997Cui}
{Cui}, W., {Springel}, V., {Yang}, X., {De Lucia}, G., \& {Borgani}, S. 2011,
  \mnras, 416, 2997

\bibitem[{{Davies} {et~al.}(2020){Davies}, {Crain}, {Oppenheimer}, \&
  {Schaye}}]{2020MNRAS.491.4462Davies}
{Davies}, J.~J., {Crain}, R.~A., {Oppenheimer}, B.~D., \& {Schaye}, J. 2020,
  \mnras, 491, 4462

\bibitem[{{Davis} {et~al.}(1985){Davis}, {Efstathiou}, {Frenk}, \&
  {White}}]{1985ApJ...292..371Davis}
{Davis}, M., {Efstathiou}, G., {Frenk}, C.~S., \& {White}, S.~D.~M. 1985, \apj,
  292, 371

\bibitem[{{Diaferio} \& {Geller}(1997)}]{1997ApJ...481..633Diaferio}
{Diaferio}, A. \& {Geller}, M.~J. 1997, \apj, 481, 633

\bibitem[{{Dolag} {et~al.}(2009){Dolag}, {Borgani}, {Murante}, \&
  {Springel}}]{2009MNRAS.399..497Dolag}
{Dolag}, K., {Borgani}, S., {Murante}, G., \& {Springel}, V. 2009, \mnras, 399,
  497

\bibitem[{{Dolag} {et~al.}(2015){Dolag}, {Gaensler}, {Beck}, \&
  {Beck}}]{2015MNRAS.451.4277Dolag}
{Dolag}, K., {Gaensler}, B.~M., {Beck}, A.~M., \& {Beck}, M.~C. 2015, \mnras,
  451, 4277

\bibitem[{{Dolag} {et~al.}(2016){Dolag}, {Komatsu}, \&
  {Sunyaev}}]{2016MNRAS.463.1797Dolag}
{Dolag}, K., {Komatsu}, E., \& {Sunyaev}, R. 2016, \mnras, 463, 1797

\bibitem[{{Duffy} {et~al.}(2010){Duffy}, {Schaye}, {Kay}, {Dalla Vecchia},
  {Battye}, \& {Booth}}]{2010MNRAS.405.2161Duffy}
{Duffy}, A.~R., {Schaye}, J., {Kay}, S.~T., {et~al.} 2010, \mnras, 405, 2161

\bibitem[{{Dvorkin} \& {Rephaeli}(2015)}]{2015MNRAS.450..896Dvorkin}
{Dvorkin}, I. \& {Rephaeli}, Y. 2015, \mnras, 450, 896

\bibitem[{{Eckert} {et~al.}(2011){Eckert}, {Molendi}, \&
  {Paltani}}]{2011A&A...526A..79Eckert}
{Eckert}, D., {Molendi}, S., \& {Paltani}, S. 2011, \aap, 526, A79

\bibitem[{{Ettori} {et~al.}(2006){Ettori}, {Dolag}, {Borgani}, \&
  {Murante}}]{2006MNRAS.365.1021Ettori}
{Ettori}, S., {Dolag}, K., {Borgani}, S., \& {Murante}, G. 2006, \mnras, 365,
  1021

\bibitem[{{Ettori} {et~al.}(2003){Ettori}, {Tozzi}, \&
  {Rosati}}]{2003A&A...398..879Ettori}
{Ettori}, S., {Tozzi}, P., \& {Rosati}, P. 2003, \aap, 398, 879

\bibitem[{{Fabjan} {et~al.}(2011){Fabjan}, {Borgani}, {Rasia}, {Bonafede},
  {Dolag}, {Murante}, \& {Tornatore}}]{2011MNRAS.416..801Fabjan}
{Fabjan}, D., {Borgani}, S., {Rasia}, E., {et~al.} 2011, \mnras, 416, 801

\bibitem[{{Fabjan} {et~al.}(2010){Fabjan}, {Borgani}, {Tornatore}, {Saro},
  {Murante}, \& {Dolag}}]{2010MNRAS.401.1670Fabjan}
{Fabjan}, D., {Borgani}, S., {Tornatore}, L., {et~al.} 2010, \mnras, 401, 1670

\bibitem[{{Farahi} {et~al.}(2019){Farahi}, {Mulroy}, {Evrard}, {Smith},
  {Finoguenov}, {Bourdin}, {Carlstrom}, {Haines}, {Marrone}, {Martino},
  {Mazzotta}, {O'Donnell}, \& {Okabe}}]{2019NatCo..10.2504Farahi}
{Farahi}, A., {Mulroy}, S.~L., {Evrard}, A.~E., {et~al.} 2019, Nature
  Communications, 10, 2504

\bibitem[{{Ferland} {et~al.}(1998){Ferland}, {Korista}, {Verner}, {Ferguson},
  {Kingdon}, \& {Verner}}]{1998PASP..110..761Ferland}
{Ferland}, G.~J., {Korista}, K.~T., {Verner}, D.~A., {et~al.} 1998, \pasp, 110,
  761

\bibitem[{{Gaspari} {et~al.}(2020){Gaspari}, {Tombesi}, \&
  {Cappi}}]{2020NatAs...4...10Gaspari}
{Gaspari}, M., {Tombesi}, F., \& {Cappi}, M. 2020, Nature Astronomy, 4, 10

\bibitem[{{Geller} {et~al.}(2013){Geller}, {Diaferio}, {Rines}, \&
  {Serra}}]{2013ApJ...764...58Geller}
{Geller}, M.~J., {Diaferio}, A., {Rines}, K.~J., \& {Serra}, A.~L. 2013, \apj,
  764, 58

\bibitem[{{Giocoli} {et~al.}(2012){Giocoli}, {Tormen}, \&
  {Sheth}}]{2012MNRAS.422..185Giocoli}
{Giocoli}, C., {Tormen}, G., \& {Sheth}, R.~K. 2012, \mnras, 422, 185

\bibitem[{{Giodini} {et~al.}(2013){Giodini}, {Lovisari}, {Pointecouteau},
  {Ettori}, {Reiprich}, \& {Hoekstra}}]{2013SSRv..177..247Giodini}
{Giodini}, S., {Lovisari}, L., {Pointecouteau}, E., {et~al.} 2013, \ssr, 177,
  247

\bibitem[{{Hirschmann} {et~al.}(2014){Hirschmann}, {Dolag}, {Saro}, {Bachmann},
  {Borgani}, \& {Burkert}}]{2014MNRAS.442.2304Hirschmann}
{Hirschmann}, M., {Dolag}, K., {Saro}, A., {et~al.} 2014, \mnras, 442, 2304

\bibitem[{{Hoekstra} {et~al.}(2015){Hoekstra}, {Herbonnet}, {Muzzin}, {Babul},
  {Mahdavi}, {Viola}, \& {Cacciato}}]{2015MNRAS.449..685Hoekstra}
{Hoekstra}, H., {Herbonnet}, R., {Muzzin}, A., {et~al.} 2015, \mnras, 449, 685

\bibitem[{{Hoekstra} {et~al.}(2012){Hoekstra}, {Mahdavi}, {Babul}, \&
  {Bildfell}}]{2012MNRAS.427.1298Hoekstra}
{Hoekstra}, H., {Mahdavi}, A., {Babul}, A., \& {Bildfell}, C. 2012, \mnras,
  427, 1298

\bibitem[{{Hudson} {et~al.}(2010){Hudson}, {Mittal}, {Reiprich}, {Nulsen},
  {Andernach}, \& {Sarazin}}]{2010A&A...513A..37Hudson}
{Hudson}, D.~S., {Mittal}, R., {Reiprich}, T.~H., {et~al.} 2010, \aap, 513, A37

\bibitem[{{Khoraminezhad} {et~al.}(2021){Khoraminezhad}, {Lazeyras}, {Angulo},
  {Hahn}, \& {Viel}}]{2021JCAP...03..023Khoraminezhad}
{Khoraminezhad}, H., {Lazeyras}, T., {Angulo}, R.~E., {Hahn}, O., \& {Viel}, M.
  2021, \jcap, 2021, 023

\bibitem[{{Komatsu} {et~al.}(2011){Komatsu}, {Smith}, {Dunkley}, {Bennett},
  {Gold}, {Hinshaw}, {Jarosik}, {Larson}, {Nolta}, {Page}, {Spergel},
  {Halpern}, {Hill}, {Kogut}, {Limon}, {Meyer}, {Odegard}, {Tucker}, {Weiland},
  {Wollack}, \& {Wright}}]{2011ApJS..192...18Komatsu}
{Komatsu}, E., {Smith}, K.~M., {Dunkley}, J., {et~al.} 2011, \apjs, 192, 18

\bibitem[{{Kravtsov} \& {Borgani}(2012)}]{2012ARA&A..50..353Kravtsov}
{Kravtsov}, A.~V. \& {Borgani}, S. 2012, \araa, 50, 353

\bibitem[{{Lima} \& {Hu}(2005)}]{2005PhRvD..72d3006Lima}
{Lima}, M. \& {Hu}, W. 2005, \prd, 72, 043006

\bibitem[{{Lu} {et~al.}(2006){Lu}, {Mo}, {Katz}, \&
  {Weinberg}}]{2006MNRAS.368.1931Lu}
{Lu}, Y., {Mo}, H.~J., {Katz}, N., \& {Weinberg}, M.~D. 2006, \mnras, 368, 1931

\bibitem[{{Ludlow} {et~al.}(2012){Ludlow}, {Navarro}, {Li}, {Angulo},
  {Boylan-Kolchin}, \& {Bett}}]{2012MNRAS.427.1322Ludlow}
{Ludlow}, A.~D., {Navarro}, J.~F., {Li}, M., {et~al.} 2012, \mnras, 427, 1322

\bibitem[{{Mantz} {et~al.}(2022){Mantz}, {Morris}, {Allen}, {Canning},
  {Baumont}, {Benson}, {Bleem}, {Ehlert}, {Floyd}, {Herbonnet}, {Kelly},
  {Liang}, {von der Linden}, {McDonald}, {Rapetti}, {Schmidt}, {Werner}, \&
  {Wright}}]{2022MNRAS.510..131Mantz}
{Mantz}, A.~B., {Morris}, R.~G., {Allen}, S.~W., {et~al.} 2022, \mnras, 510,
  131

\bibitem[{{Maughan} {et~al.}(2016){Maughan}, {Giles}, {Rines}, {Diaferio},
  {Geller}, {Van Der Pyl}, \& {Bonamente}}]{2016MNRAS.461.4182Maughan}
{Maughan}, B.~J., {Giles}, P.~A., {Rines}, K.~J., {et~al.} 2016, \mnras, 461,
  4182

\bibitem[{{Melchior} {et~al.}(2015){Melchior}, {Suchyta}, {Huff}, {Hirsch},
  {Kacprzak}, {Rykoff}, {Gruen}, {Armstrong}, {Bacon}, {Bechtol}, {Bernstein},
  {Bridle}, {Clampitt}, {Honscheid}, {Jain}, {Jouvel}, {Krause}, {Lin},
  {MacCrann}, {Patton}, {Plazas}, {Rowe}, {Vikram}, {Wilcox}, {Young}, {Zuntz},
  {Abbott}, {Abdalla}, {Allam}, {Banerji}, {Bernstein}, {Bernstein}, {Bertin},
  {Buckley-Geer}, {Burke}, {Castander}, {da Costa}, {Cunha}, {Depoy}, {Desai},
  {Diehl}, {Doel}, {Estrada}, {Evrard}, {Neto}, {Fernandez}, {Finley},
  {Flaugher}, {Frieman}, {Gaztanaga}, {Gerdes}, {Gruendl}, {Gutierrez},
  {Jarvis}, {Karliner}, {Kent}, {Kuehn}, {Kuropatkin}, {Lahav}, {Maia},
  {Makler}, {Marriner}, {Marshall}, {Merritt}, {Miller}, {Miquel}, {Mohr},
  {Neilsen}, {Nichol}, {Nord}, {Reil}, {Roe}, {Roodman}, {Sako}, {Sanchez},
  {Santiago}, {Schindler}, {Schubnell}, {Sevilla-Noarbe}, {Sheldon}, {Smith},
  {Soares-Santos}, {Swanson}, {Sypniewski}, {Tarle}, {Thaler}, {Thomas},
  {Tucker}, {Walker}, {Wechsler}, {Weller}, \&
  {Wester}}]{2015MNRAS.449.2219Melchior}
{Melchior}, P., {Suchyta}, E., {Huff}, E., {et~al.} 2015, \mnras, 449, 2219

\bibitem[{{Naderi} {et~al.}(2015){Naderi}, {Malekjani}, \&
  {Pace}}]{2015MNRAS.447.1873Naderi}
{Naderi}, T., {Malekjani}, M., \& {Pace}, F. 2015, \mnras, 447, 1873

\bibitem[{{Navarro} {et~al.}(1997){Navarro}, {Frenk}, \&
  {White}}]{1997ApJ...490..493Navarro}
{Navarro}, J.~F., {Frenk}, C.~S., \& {White}, S. D.~M. 1997, \apj, 490, 493

\bibitem[{{Okabe} {et~al.}(2010){Okabe}, {Zhang}, {Finoguenov}, {Takada},
  {Smith}, {Umetsu}, \& {Futamase}}]{2010ApJ...721..875Okabe}
{Okabe}, N., {Zhang}, Y.~Y., {Finoguenov}, A., {et~al.} 2010, \apj, 721, 875

\bibitem[{{Ostriker} {et~al.}(2005){Ostriker}, {Bode}, \&
  {Babul}}]{2005ApJ...634..964Ostriker}
{Ostriker}, J.~P., {Bode}, P., \& {Babul}, A. 2005, \apj, 634, 964

\bibitem[{{Pacaud} {et~al.}(2007){Pacaud}, {Pierre}, {Adami}, {Altieri},
  {Andreon}, {Chiappetti}, {Detal}, {Duc}, {Galaz}, {Gueguen}, {Le F{\`e}vre},
  {Hertling}, {Libbrecht}, {Melin}, {Ponman}, {Quintana}, {Refregier},
  {Sprimont}, {Surdej}, {Valtchanov}, {Willis}, {Alloin}, {Birkinshaw},
  {Bremer}, {Garcet}, {Jean}, {Jones}, {Le F{\`e}vre}, {Maccagni}, {Mazure},
  {Proust}, {R{\"o}ttgering}, \& {Trinchieri}}]{2007MNRAS.382.1289Pacaud}
{Pacaud}, F., {Pierre}, M., {Adami}, C., {et~al.} 2007, \mnras, 382, 1289

\bibitem[{{Planelles} {et~al.}(2013){Planelles}, {Borgani}, {Dolag}, {Ettori},
  {Fabjan}, {Murante}, \& {Tornatore}}]{2013MNRAS.431.1487Planelles}
{Planelles}, S., {Borgani}, S., {Dolag}, K., {et~al.} 2013, \mnras, 431, 1487

\bibitem[{{Poole} {et~al.}(2007){Poole}, {Babul}, {McCarthy}, {Fardal},
  {Bildfell}, {Quinn}, \& {Mahdavi}}]{2007MNRAS.380..437Poole}
{Poole}, G.~B., {Babul}, A., {McCarthy}, I.~G., {et~al.} 2007, \mnras, 380, 437

\bibitem[{{Pratt} {et~al.}(2009){Pratt}, {Croston}, {Arnaud}, \&
  {B{\"o}hringer}}]{2009A&A...498..361Pratt}
{Pratt}, G.~W., {Croston}, J.~H., {Arnaud}, M., \& {B{\"o}hringer}, H. 2009,
  \aap, 498, 361

\bibitem[{{Puddu} \& {Andreon}(2022)}]{2022MNRAS.511.2968Puddu}
{Puddu}, E. \& {Andreon}, S. 2022, \mnras, 511, 2968

\bibitem[{{Ragagnin} {et~al.}(2017){Ragagnin}, {Dolag}, {Biffi}, {Cadolle Bel},
  {Hammer}, {Krukau}, {Petkova}, \& {Steinborn}}]{2017A&C....20...52Ragagnin}
{Ragagnin}, A., {Dolag}, K., {Biffi}, V., {et~al.} 2017, Astronomy and
  Computing, 20, 52

\bibitem[{{Ragagnin} {et~al.}(2019){Ragagnin}, {Dolag}, {Moscardini},
  {Biviano}, \& {D'Onofrio}}]{2019MNRAS.486.4001Ragagnin}
{Ragagnin}, A., {Dolag}, K., {Moscardini}, L., {Biviano}, A., \& {D'Onofrio},
  M. 2019, \mnras, 486, 4001

\bibitem[{{Ragagnin} {et~al.}(2021){Ragagnin}, {Saro}, {Singh}, \&
  {Dolag}}]{2021MNRAS.500.5056Ragagnin}
{Ragagnin}, A., {Saro}, A., {Singh}, P., \& {Dolag}, K. 2021, \mnras, 500, 5056

\bibitem[{{Ragagnin} {et~al.}(2016){Ragagnin}, {Tchipev}, {Bader}, {Dolag}, \&
  {Hammer}}]{2016pcre.conf..411Ragagnin}
{Ragagnin}, A., {Tchipev}, N., {Bader}, M., {Dolag}, K., \& {Hammer}, N.~J.
  2016, in Advances in Parallel Computing, Volume 27: Parallel Computing: On
  the Road to Exascale, Edited by Gerhard R. Joubert, Hugh Leather, Mark
  Parsons, Frans Peters, Mark Sawyer. IOP Ebook, ISBN: 978-1-61499-621-7, pages
  411-420

\bibitem[{{Ragone-Figueroa} {et~al.}(2010){Ragone-Figueroa}, {Plionis},
  {Merch{\'a}n}, {Gottl{\"o}ber}, \&
  {Yepes}}]{2010MNRAS.407..581RagoneFigueroa}
{Ragone-Figueroa}, C., {Plionis}, M., {Merch{\'a}n}, M., {Gottl{\"o}ber}, S.,
  \& {Yepes}, G. 2010, \mnras, 407, 581

\bibitem[{{Remus} {et~al.}(2017){Remus}, {Dolag}, {Naab}, {Burkert},
  {Hirschmann}, {Hoffmann}, \& {Johansson}}]{2017MNRAS.464.3742Remus}
{Remus}, R.-S., {Dolag}, K., {Naab}, T., {et~al.} 2017, \mnras, 464, 3742

\bibitem[{{Richardson} \& {Corasaniti}(2021)}]{2021arXiv211204926Richardson}
{Richardson}, T.~R.~G. \& {Corasaniti}, P.~S. 2021, arXiv e-prints,
  arXiv:2112.04926

\bibitem[{{Saro} {et~al.}(2014){Saro}, {Liu}, {Mohr}, {Aird}, {Ashby},
  {Bayliss}, {Benson}, {Bleem}, {Bocquet}, {Brodwin}, {Carlstrom}, {Chang},
  {Chiu}, {Cho}, {Clocchiatti}, {Crawford}, {Crites}, {de Haan}, {Desai},
  {Dietrich}, {Dobbs}, {Dolag}, {Dudley}, {Foley}, {Gangkofner}, {George},
  {Gladders}, {Gonzalez}, {Halverson}, {Hennig}, {Hlavacek-Larrondo},
  {Holzapfel}, {Hrubes}, {Jones}, {Keisler}, {Lee}, {Leitch}, {Lueker},
  {Luong-Van}, {Mantz}, {Marrone}, {McDonald}, {McMahon}, {Mehl}, {Meyer},
  {Mocanu}, {Montroy}, {Murray}, {Nurgaliev}, {Padin}, {Patej}, {Pryke},
  {Reichardt}, {Rest}, {Ruel}, {Ruhl}, {Saliwanchik}, {Sayre}, {Schaffer},
  {Shirokoff}, {Spieler}, {Stalder}, {Staniszewski}, {Stark}, {Story}, {van
  Engelen}, {Vanderlinde}, {Vieira}, {Vikhlinin}, {Williamson}, {Zahn}, \&
  {Zenteno}}]{2014MNRAS.440.2610Saro}
{Saro}, A., {Liu}, J., {Mohr}, J.~J., {et~al.} 2014, \mnras, 440, 2610

\bibitem[{{Schrabback} {et~al.}(2021){Schrabback}, {Bocquet}, {Sommer},
  {Zohren}, {van den Busch}, {Hern{\'a}ndez-Mart{\'\i}n}, {Hoekstra}, {Raihan},
  {Schirmer}, {Applegate}, {Bayliss}, {Benson}, {Bleem}, {Dietrich}, {Floyd},
  {Hilbert}, {Hlavacek-Larrondo}, {McDonald}, {Saro}, {Stark}, \&
  {Weissgerber}}]{2021MNRAS.505.3923Schrabback}
{Schrabback}, T., {Bocquet}, S., {Sommer}, M., {et~al.} 2021, \mnras, 505, 3923

\bibitem[{{Singh} {et~al.}(2020){Singh}, {Saro}, {Costanzi}, \&
  {Dolag}}]{2020MNRAS.494.3728Singh}
{Singh}, P., {Saro}, A., {Costanzi}, M., \& {Dolag}, K. 2020, \mnras, 494, 3728

\bibitem[{{Smith} {et~al.}(2001){Smith}, {Brickhouse}, {Liedahl}, \&
  {Raymond}}]{2001ApJ...556L..91Smith}
{Smith}, R.~K., {Brickhouse}, N.~S., {Liedahl}, D.~A., \& {Raymond}, J.~C.
  2001, \apjl, 556, L91

\bibitem[{{Springel}(2005)}]{2005MNRAS.364.1105Springel}
{Springel}, V. 2005, \mnras, 364, 1105

\bibitem[{{Springel} {et~al.}(2005{\natexlab{a}}){Springel}, {Di Matteo}, \&
  {Hernquist}}]{2005MNRAS.361..776Springel}
{Springel}, V., {Di Matteo}, T., \& {Hernquist}, L. 2005{\natexlab{a}}, \mnras,
  361, 776

\bibitem[{{Springel} {et~al.}(2005{\natexlab{b}}){Springel}, {White},
  {Jenkins}, {Frenk}, {Yoshida}, {Gao}, {Navarro}, {Thacker}, {Croton},
  {Helly}, {Peacock}, {Cole}, {Thomas}, {Couchman}, {Evrard}, {Colberg}, \&
  {Pearce}}]{2005Natur.435..629Springel}
{Springel}, V., {White}, S.~D.~M., {Jenkins}, A., {et~al.} 2005{\natexlab{b}},
  \nat, 435, 629

\bibitem[{{Springel} {et~al.}(2001){Springel}, {White}, {Tormen}, \&
  {Kauffmann}}]{2001MNRAS.328..726Springel}
{Springel}, V., {White}, S.~D.~M., {Tormen}, G., \& {Kauffmann}, G. 2001,
  \mnras, 328, 726

\bibitem[{{Stanek} {et~al.}(2010){Stanek}, {Rasia}, {Evrard}, {Pearce}, \&
  {Gazzola}}]{2010ApJ...715.1508Stanek}
{Stanek}, R., {Rasia}, E., {Evrard}, A.~E., {Pearce}, F., \& {Gazzola}, L.
  2010, \apj, 715, 1508

\bibitem[{{Steinborn} {et~al.}(2016){Steinborn}, {Dolag}, {Comerford},
  {Hirschmann}, {Remus}, \& {Teklu}}]{2016MNRAS.458.1013Steinborn}
{Steinborn}, L.~K., {Dolag}, K., {Comerford}, J.~M., {et~al.} 2016, \mnras,
  458, 1013

\bibitem[{{Steinborn} {et~al.}(2015){Steinborn}, {Dolag}, {Hirschmann},
  {Prieto}, \& {Remus}}]{2015MNRAS.448.1504Steinborn}
{Steinborn}, L.~K., {Dolag}, K., {Hirschmann}, M., {Prieto}, M.~A., \& {Remus},
  R.-S. 2015, \mnras, 448, 1504

\bibitem[{{Stern} {et~al.}(2019){Stern}, {Dietrich}, {Bocquet}, {Applegate},
  {Mohr}, {Bridle}, {Carrasco Kind}, {Gruen}, {Jarvis}, {Kacprzak}, {Saro},
  {Sheldon}, {Troxel}, {Zuntz}, {Benson}, {Capasso}, {Chiu}, {Desai},
  {Rapetti}, {Reichardt}, {Saliwanchik}, {Schrabback}, {Gupta}, {Abbott},
  {Abdalla}, {Avila}, {Bertin}, {Brooks}, {Burke}, {Carnero Rosell},
  {Carretero}, {Castander}, {D'Andrea}, {da Costa}, {Davis}, {De Vicente},
  {Diehl}, {Doel}, {Estrada}, {Evrard}, {Flaugher}, {Fosalba}, {Frieman},
  {Garc{\'\i}a-Bellido}, {Gaztanaga}, {Gruendl}, {Gschwend}, {Gutierrez},
  {Hollowood}, {Jeltema}, {Kirk}, {Kuehn}, {Kuropatkin}, {Lahav}, {Lima},
  {Maia}, {March}, {Melchior}, {Menanteau}, {Miquel}, {Plazas}, {Romer},
  {Sanchez}, {Schindler}, {Schubnell}, {Sevilla-Noarbe}, {Smith}, {Smith},
  {Sobreira}, {Suchyta}, {Swanson}, {Tarle}, {Walker}, {DES Collaboration}, \&
  {SPT Collaboration}}]{2019MNRAS.485...69Stern}
{Stern}, C., {Dietrich}, J.~P., {Bocquet}, S., {et~al.} 2019, \mnras, 485, 69

\bibitem[{{Sun} {et~al.}(2009){Sun}, {Voit}, {Donahue}, {Jones}, {Forman}, \&
  {Vikhlinin}}]{2009ApJ...693.1142Sun}
{Sun}, M., {Voit}, G.~M., {Donahue}, M., {et~al.} 2009, \apj, 693, 1142

\bibitem[{{Taffoni} {et~al.}(2020){Taffoni}, {Becciani}, {Garilli}, {Maggio},
  {Pasian}, {Umana}, {Smareglia}, \& {Vitello}}]{2020ASPC..527..307Taffoni}
{Taffoni}, G., {Becciani}, U., {Garilli}, B., {et~al.} 2020, in Astronomical
  Society of the Pacific Conference Series, Vol. 527, Astronomical Society of
  the Pacific Conference Series, ed. R.~{Pizzo}, E.~R. {Deul}, J.~D. {Mol},
  J.~{de Plaa}, \& H.~{Verkouter}, 307

\bibitem[{{Teklu} {et~al.}(2015){Teklu}, {Remus}, {Dolag}, {Beck}, {Burkert},
  {Schmidt}, {Schulze}, \& {Steinborn}}]{2015ApJ...812...29Teklu}
{Teklu}, A.~F., {Remus}, R.-S., {Dolag}, K., {et~al.} 2015, \apj, 812, 29

\bibitem[{Tornatore {et~al.}(2007)Tornatore, Borgani, Dolag, \&
  Matteucci}]{2007MNRAS.382.1050Tornatore}
Tornatore, L., Borgani, S., Dolag, K., \& Matteucci, F. 2007, Monthly Notices
  of the Royal Astronomical Society, 382, 1050

\bibitem[{{Torri} {et~al.}(2004){Torri}, {Meneghetti}, {Bartelmann},
  {Moscardini}, {Rasia}, \& {Tormen}}]{2004MNRAS.349..476Torri}
{Torri}, E., {Meneghetti}, M., {Bartelmann}, M., {et~al.} 2004, \mnras, 349,
  476

\bibitem[{{Truemper}(1993)}]{1993Sci...260.1769Truemper}
{Truemper}, J. 1993, Science, 260, 1769

\bibitem[{{Truong} {et~al.}(2018){Truong}, {Rasia}, {Mazzotta}, {Planelles},
  {Biffi}, {Fabjan}, {Beck}, {Borgani}, {Dolag}, {Gaspari}, {Granato},
  {Murante}, {Ragone-Figueroa}, \& {Steinborn}}]{2018MNRAS.474.4089Truong}
{Truong}, N., {Rasia}, E., {Mazzotta}, P., {et~al.} 2018, \mnras, 474, 4089

\bibitem[{{Vall{\'e}s-P{\'e}rez} {et~al.}(2020){Vall{\'e}s-P{\'e}rez},
  {Planelles}, \& {Quilis}}]{2020MNRAS.499.2303VallesPerez}
{Vall{\'e}s-P{\'e}rez}, D., {Planelles}, S., \& {Quilis}, V. 2020, \mnras, 499,
  2303

\bibitem[{{van de Sande} {et~al.}(2019){van de Sande}, {Lagos}, {Welker},
  {Bland-Hawthorn}, {Schulze}, {Remus}, {Bah{\'e}}, {Brough}, {Bryant},
  {Cortese}, {Croom}, {Devriendt}, {Dubois}, {Goodwin}, {Konstantopoulos},
  {Lawrence}, {Medling}, {Pichon}, {Richards}, {Sanchez}, {Scott}, \&
  {Sweet}}]{2019MNRAS.484..869VanDeSande}
{van de Sande}, J., {Lagos}, C.~D.~P., {Welker}, C., {et~al.} 2019, \mnras,
  484, 869

\bibitem[{{Velliscig} {et~al.}(2014){Velliscig}, {van Daalen}, {Schaye},
  {McCarthy}, {Cacciato}, {Le Brun}, \& {Dalla
  Vecchia}}]{2014MNRAS.442.2641Velliscig}
{Velliscig}, M., {van Daalen}, M.~P., {Schaye}, J., {et~al.} 2014, \mnras, 442,
  2641

\bibitem[{{Vikhlinin} {et~al.}(2006){Vikhlinin}, {Kravtsov}, {Forman}, {Jones},
  {Markevitch}, {Murray}, \& {Van Speybroeck}}]{2006ApJ...640..691Vikhlinin}
{Vikhlinin}, A., {Kravtsov}, A., {Forman}, W., {et~al.} 2006, \apj, 640, 691

\bibitem[{{Wang} {et~al.}(2020){Wang}, {Mao}, {Zentner}, {Lange}, {van den
  Bosch}, \& {Wechsler}}]{2020MNRAS.498.4450Wang}
{Wang}, K., {Mao}, Y.-Y., {Zentner}, A.~R., {et~al.} 2020, \mnras, 498, 4450

\bibitem[{{Wechsler} {et~al.}(2002){Wechsler}, {Bullock}, {Primack},
  {Kravtsov}, \& {Dekel}}]{2002ApJ...568...52Wechsler}
{Wechsler}, R.~H., {Bullock}, J.~S., {Primack}, J.~R., {Kravtsov}, A.~V., \&
  {Dekel}, A. 2002, \apj, 568, 52

\bibitem[{{Wechsler} \& {Tinker}(2018)}]{2018ARA&A..56..435Wechsler}
{Wechsler}, R.~H. \& {Tinker}, J.~L. 2018, \araa, 56, 435

\bibitem[{{Xu} {et~al.}(2018){Xu}, {Ramos-Ceja}, {Pacaud}, {Reiprich}, \&
  {Erben}}]{2018A&A...619A.162Xu}
{Xu}, W., {Ramos-Ceja}, M.~E., {Pacaud}, F., {Reiprich}, T.~H., \& {Erben}, T.
  2018, \aap, 619, A162

\bibitem[{{Zhao} {et~al.}(2003){Zhao}, {Mo}, {Jing}, \&
  {B{\"o}rner}}]{2003MNRAS.339...12Zhao}
{Zhao}, D.~H., {Mo}, H.~J., {Jing}, Y.~P., \& {B{\"o}rner}, G. 2003, \mnras,
  339, 12

\end{thebibliography}
 
\end{document}